\documentclass[11pt]{amsart}
\usepackage{amsfonts}
\usepackage{amsmath,amsthm}
\usepackage{bbm}

\usepackage{hyperref}
\usepackage[usenames,dvipsnames]{color}
\hypersetup{colorlinks,
	citecolor=Blue,
	linkcolor=Blue,
	urlcolor=Blue}

\usepackage{latexsym}
\usepackage{latexsym}
\usepackage{array}
\usepackage{amssymb}
\usepackage{enumerate}
\usepackage{comment}
\usepackage{graphicx}
\usepackage{tikz}
\usepackage{color}
\usepackage{bbm}

\DeclareFontFamily{U}{wncy}{}
\DeclareFontShape{U}{wncy}{m}{n}{<->wncyr10}{}
\DeclareSymbolFont{mcy}{U}{wncy}{m}{n}
\DeclareMathSymbol{\Sh}{\mathord}{mcy}{"58}

\usepackage[english]{babel}
\usepackage{fdsymbol}
\usepackage{color}
\usepackage[latin1]{inputenc}

\numberwithin{equation}{section}

\theoremstyle{plain}
\newtheorem{theorem}{Theorem}[section]
\newtheorem*{theorem*}{Theorem}
\newtheorem{lemma}[theorem]{Lemma}
\newtheorem{proposition}[theorem]{Proposition}
\newtheorem{corollary}[theorem]{Corollary}

\newtheorem{conjecture}[theorem]{Conjecture}

\theoremstyle{remark}
\newtheorem{remark}[theorem]{Remark}

\newtheorem*{lem*}{Lemma}
\newtheorem*{sublem*}{Sublemma}
\newtheorem*{remark*}{Remark}
\newtheorem*{NB*}{NB}

\newcommand{\PP}{ \mathbb{P}\,}

\newcommand{\supp}{\mathop{\rm supp}\nolimits}

\newcommand{\R}{ \mathbb{R} }

\newcommand{\C}{ \mathbb{C} }
\newcommand{\Z}{ \mathbb{Z} }
\newcommand{\N}{ \mathbb{N} }

\newcommand{\T}{ \mathbb{T} }

\newcommand{\mI}{\mathbb{I}}

\newcommand{\fz}{\mathfrak{z}}

\newcommand{\Del}{\Delta}

\newcommand{\cA}{ \mathcal{A} }
\newcommand{\cB}{ \mathcal{B} }
\newcommand{\cC}{ \mathcal{C} }
\newcommand{\cP}{ \mathcal{P} }

\newcommand{\cY}{ \mathcal{Y} }
\newcommand{\cL}{ \mathcal{L} }
\newcommand{\cR}{ \mathcal{R} }

\newcommand{\cK}{ \mathcal{K} }
\newcommand{\cF}{ \mathcal{F} }

\newcommand{\EE}{ {\mathbb E}}
\newcommand{\F}{ \mathcal{F} }

\newcommand{\cZ}{ \mathcal{Z} }
\newcommand{\cM}{ \mathcal{M} }
\newcommand{\cN}{ \mathcal{N} }

\renewcommand{\P}{ \mathcal{P} }

\newcommand{\cT}{ \mathcal{T} }
\newcommand{\cW}{ \mathcal{W} }
\newcommand{\cJ}{ \mathcal{J} }

\newcommand{\om}{ \omega }

\newcommand{\Om}{ \Omega }

\newcommand{\ga}{\gamma }

\newcommand{\s}{ \sigma }

\newcommand{\ka}{ \kappa }

\renewcommand{\phi}{ \varphi }

\newcommand{\eps}{\varepsilon}

\newcommand{\de}{ \delta }
\newcommand{\al}{ \alpha }
\newcommand{\zz}{\mathfrak z}
\newcommand{\fm}{\mathfrak m}

\newcommand{\la}{\lambda}

\newcommand{\Id}{\operatorname{Id}}

\newcommand{\be}{\begin{equation}}
	\newcommand{\ee}{\end{equation}}
\newcommand{\ben}{\begin{equation*}}
	\newcommand{\een}{\end{equation*}}

\newcommand{\ov}{ \overline }

\newcommand{\lan}{ \langle }

\newcommand{\ran}{ \rangle}

\newcommand{\p}{ \partial}

\newcommand{\e}{ \text{e} }

\newcommand{\wt}{ \widetilde }

\newcommand{\lbl}{\label}
\newcommand{\non}{\nonumber}
\newcommand{\qu}{\quad}
\newcommand{\qmb}{\quad\mbox}
\newcommand{\qnd}{\qmb{and}\qu}

\newcommand{\Ga}{\Gamma}

\newcommand{\sm}{\setminus}

\newcommand{\sign}{\operatorname{sign}}

\newcommand{\ssk}{\smallskip}
\newcommand{\msk}{\medskip}
\newcommand{\bsk}{\bigskip}

\newcommand{\ds}{\displaystyle}

\newcommand{\bm}{\boldsymbol}
\newcommand{\bbeta}{\boldsymbol{\eta}}
\newcommand{\bxi}{\boldsymbol{\xi}}
\newcommand{\bsi}{\boldsymbol{\sigma}}

\title [kinetic approximation for equations of discrete turbulence]
{Kinetic approximation for equations of discrete turbulence in the subcritical case }

\author{Andrey  Dymov}
\address{Steklov Mathematical Institute of RAS, Moscow, Russia
		\& National Research University Higher School of
	Economics, Moscow, Russia
	\& Skolkovo Institute of Science and Technology, Skolkovo, Russia 
} \email{dymov@mi-ras.ru}
\begin{document}
	\begin{abstract}
		We consider a damped/driven cubic NLS equation on a torus under the limit when first the amplitude of solutions goes to zero and then the period of the torus goes to infinity. We suggest another proof of the kinetic approximation for the energy spectrum under a subcritical scaling, extending to the exact solutions result obtained in [Dymov, Kuksin, Maiocchi, Vladuts '2023] for quasisolutions which were defined as the second order truncations of decompositions for the solutions in amplitude. The proof does not involve Feynman diagrams, instead relying on a robust inductive analysis of cumulants.
		\end{abstract}

	\maketitle
	\date{}
	
		\bigskip
	\tableofcontents
	
	\section{Introduction}

\subsection{The setting}	
The wave turbulence theory (WT) has been intensively developing in physical works since the 1960's (\cite{ZLF92, Naz11, NR}).
It can be viewed as a kinetic theory of dispersive nonlinear wave systems, going back to the famous R.~Peierls' kinetic theory from 1929 of the heat conduction in crystals \cite{Pei}. On the other hand it can be considered from the perspective of the strong, hydrodynamic turbulence since the WT theory predicts energy cascades similar to the one discovered by A.~Kolmogorov (see \cite{Bed} in addition to the references above).  

From the mathematical point of view WT is a heuristic approach for studying small amplitude solutions to nonlinear dispersive wave systems in hydrodynamic limit. 
There is strong long standing interest in the community of mathematical physicists to the problem of rigorous justification of the WT's predictions but progress in studying this question was achieved only recently, see Section~\ref{s:res_dissc}. 

Mathematical study of the questions addressed by the WT requires stochastization of a  deterministic wave system. In particular, one can assume that the initial data is random or,  following Zakharov-L'vov \cite{ZL75}, add to the equations of motion a stochastic perturbation together with damping, balancing the latter.  The second choice provides a rigorous basis for studying the regime, often discussed by members of the WT community, when "energy is pumped  to low modes and dissipated in high modes", in which they perform numerical and empirical experiments. 

In the present paper we continue to study the Zakharov-L'vov stochastic model for WT, started in \cite{DK1, DK2, DKMV, HG}. More specifically, we consider a damped/driven cubic NLS equation on a $d$-dimensional torus  $\T^d_L = \R^d/(L\Z^d)$, $d\ge 2$, of period $L\ge 1$,
\begin{equation}\label{NLS}
	\p_t u +i  \Delta u  - i\la\,\big( |u|^2 -2\|u\|^2\big)u =  -\nu\frak A(u)    + \sqrt\nu\p_t \eta^\omega(t,x),
\end{equation}
where $u=u(t,x)$, $x\in \T^d_L$,  $ \Delta=(2\pi)^{-2}\sum_{j=1}^d ({{\partial}}^2 / {{\partial}} x_j^2)$ is the Laplace operator, $\la$ and $\nu$ are positive small parameters and
\be\lbl{norm_L_2}
\|u\|^2 :=
L^{-d}\int_{{{\mathbb T}}^d_L} |u(x)|^2\,dx
\ee
is the normalized over the volume $L_2$-norm.  
The modification of the cubic nonlinearity $i|u|^2u$ in \eqref{NLS} by the term  $2i\|u\|^2 u$ is common in the WT. It keeps the main features of the standard cubic NLS equation, reducing some non-critical technicalities; see introduction to the \cite{DK1}. 

The terms  $-\nu\frak A(u)$ and  $\sqrt\nu\p_t\eta^\om$ represent damping and  white noise stochastic perturbation, and are defined below. A possible choice for the linear operator $\frak A$ is $\frak A(u) = (1-\Delta)^{r_*}$ with $r_*>0$.
The scaling $\nu$ and $\sqrt\nu$ of the damping and noise is standard since it keeps balance of the two terms as $\nu\to 0$.
It is natural to require that $\nu\ll\la$ since then the r.h.s. of eq. \eqref{NLS} is a small perturbation of the deterministic NLS equation in the l.h.s.; 
later we will assume a kind of this relation.

Let us define the Fourier series for a functions $u(x)$ on $\T^d_L$ as
\begin{equation}\lbl{Fourier-def}
	u(x)= L^{-d/2} \,\sum\limits_{s\in{{\mathbb Z}}^d_L} \hat u_s e^{2\pi  i s\cdot x}, 
	\qquad\qu{{\mathbb Z}}^d_L := L^{-1} {{\mathbb Z}}^d\,.
\end{equation}
 The Fourier coefficients $\hat u_s$ are given by the Fourier transform
\be\non
\hat u=\cF(u), \quad \hat u_s = L^{-d/2}\,\int_{\T^d_L} u(x) e^{-2\pi i s\cdot x}\,dx, \qquad s\in \Z_L^d. 
\ee
The scaling $L^{-d/2}$ of series \eqref{Fourier-def} is convenient since, 
as we will show, it
 leads to the Fourier coefficients $\hat u_s$ of solutions to \eqref{NLS} of size $O_s(1)$ uniformly in (appropriately scaled) parameters $\nu,\la,L$.
The  Parseval identity for the norm \eqref{norm_L_2} reads
\[
\|u\|^2 = L^{-d}\,\sum\limits_{s\in\Z^d_L} |\hat u_s|^2.
\] 
 We define the damping linear operator $\frak A$ as $\frak A = \ga^0(-\Delta)$, where
  $\gamma^0$ is a  real smooth  increasing function of a positive argument, 
 satisfying 
 \be\label{gamma}
 \gamma^0\ge1 \;\; \text{and}\;\; 
 c(1+ y)^{r_*} \le \gamma^0(y)  \le  C(1+ y)^{r_*}\qquad
 \forall\, y>0\,,
 \ee
 for some positive constants $r_*, c, C$. We also  assume that 
 all derivatives of  $\ga^0$ have at most polynomial growths at infinity.
 Explicitly,
 \be\lbl{diss_op}
\frak A(u(x)) = L^{-d/2}\sum\limits_{s\in \Z^d_L} \ga_s \hat u_s e^{2\pi is\cdot x},  \quad \ga_s := \ga^0\big(|s|^2\big),
\ee
where $\hat u = \cF(u)$ and $|s|$ is the Euclidean norm of a vector $s\in\Z^d_L$.
E.g., for $\ga_s=(1+|s|^2)^{r_*}$ we have $\frak A = (1-\Delta)^{r_*}$.

The random process  $\eta^\omega$ is defined  via the Fourier series
$$
\eta^\omega(t,x)=
L^{-d/2} \, \sum\limits_{s\in\Z^d_L} b(s) \beta^\omega_s(t) e^{2\pi  i s\cdot x},
$$
where
$\{\beta_s(t), s\in{{\mathbb Z}}^d_L\}$ are standard independent complex\footnote{
	I.e. $\beta_s = \beta_s^1 + i\beta_s^2$, where $\{\beta_s^j, s\in{{\mathbb Z}}^d_L, j=1,2  \}$ are
	standard independent real Wiener processes.} Wiener processes  and  
$b(s)$
is a 
	positive  Schwartz  function on ${{\mathbb R}}^d$. 

	We work on long time intervals $t\sim\nu^{-1}$, so that the stochastic perturbation and damping give input of order one to a solution of \eqref{NLS}. Then it is convenient to write  equation \eqref{NLS} in the slow time $\tau = \nu t$, $\tau\sim 1$:
	\begin{align}\non
			\dot u +i \nu^{-1} \Delta u  - i\la\nu^{-1}\,\big( |u|^2 -2\|u\|^2\big)u &= -\frak A(u)    + \dot \eta^\omega(\tau,x),\\ \label{NLS_st}
			\eta^\omega(\tau,x)&=
			L^{-d/2} \, {\sum}_{s\in\Z^d_L} b(s) \beta^\omega_s(\tau) e^{2\pi  i s\cdot x}\,.
	\end{align}
	Here the upper-dot denotes the derivative $\p_\tau$ in the slow time
	and $\{\beta_s(\tau), \, s\in\Z^d_L\}$ is another set of standard independent complex Wiener processes.  

	We regard solutions to equation \eqref{NLS_st} as random processes in the space $L_2(\T^d_L)$, provided with the norm \eqref{norm_L_2}. It is known that equation \eqref{NLS_st} is well posed and mixing
\footnote{I.e. it has a unique stationary measure $\mu$ and any its solution with deterministic initial data converges to $\mu$ in distribution when $\tau\to\infty.$}
once $r_*$ is sufficiently large, see Section~\ref{s:dt} and \cite[Section 1.2]{DKMV}. 
Moreover, applying the Ito formula it is straightforward to see (\cite[Section 1]{DK1}) that "averaged energy per volume", given by the expectation $\EE \|u(\tau)\|^2$, stays of size one uniformly in time $\tau>0$  and the parameters $\nu,  \la, L$,  provided that   $\EE \|u(0)\|^2$  does. 

The objective of WT is to study solutions  of \eqref{NLS_st} and their statistics when 
\be\lbl{limmm}
\nu,\; \la\to 0 \qmb{and}\qu L\to\infty.
\ee
More specifically, in the centre of attention of the WT community is the behaviour under limit \eqref{limmm} of the  {\it energy spectrum}
\be\lbl{es}
\frak n_s(\tau):= \EE|\hat u_s(\tau)|^2, \qquad s\in\Z^d_L, \qu \tau\ge 0,\qquad \hat u = \cF(u),
\ee
that describes the share of the  $L_2$-norm (preserved by the deterministic equation \eqref{NLS}$\big|_{\nu=0}$) over specific Fourier frequencies.
One expects (see \cite{ZLF92, Naz11}) that under limit \eqref{limmm} behaviour of the energy spectrum is governed by a nonlinear kinetic equation called the {\it wave kinetic equation} (WKE), which might be damped and driven   since equation \eqref{NLS_st} is.  
Note that the WT theory does not postulate any relation for the parameters $\nu,\la$ and $L$.

In this paper we continue the study started in \cite{DKMV}, specifying the limit \eqref{limmm} as 
\be\lbl{lim}
\mbox{first}\qu\nu \to 0 \qmb{and then}\qu  L\to\infty,
\ee
while $\la=\la(\nu,L)$ is scaled appropriately.
We establish the expected approximation of the energy spectrum $\frak n_s(\tau)$ by a solution to the WKE in a {\it subcritical} scaling of the parameter $\la$, defined below in \eqref{la_sc}, \eqref{subcr}. 
The obtained WKE \eqref{WKE} turns out to be {\it modified} (in particular, non-autonomous) although it is similar to that arising in physical papers.  

In recent years a number of works was published, studying the behaviour of energy spectrum \eqref{es} of solutions 
to eq.  \eqref{NLS_st} and eq. \eqref{NLS}$\big|_{\nu=0}$ with random initial data, under limit \eqref{limmm} with various scalings, different from \eqref{lim}. See Section~\ref{s:res_dissc} for a short review.  Due to our knowledge all existing results rely on a powerful but very sophisticated method of Feynman diagrams decomposition. In our work we do not use the Feynman diagrams, relying instead on a relatively simple and robust inductive analysis of cumulants. 
See Section~\ref{s:Fd-cum} for a discussion.

\subsection{Results}
In this section we briefly discuss our results while their detailed statement together with schemes of the proofs are given in Section~\ref{s:res_det}.

In the main body of the paper we focus on the case $d\ge 3$. The modifications, required for $d=2$, are explained in Section~\ref{s:d=2}.
	To simplify the presentation we assume that initially the system is at rest, providing eq. \eqref{NLS_st} with the initial conditions
\be\lbl{in_data}
u(x,0) = 0.
\ee
 In  \cite{DKMV} we show that if $d\ge 3$, under limit \eqref{lim} the energy spectrum $\frak n_s(\tau)$ behaves non-trivially
\footnote{I.e. stays finite and behaves differently from $\EE |\hat u_s^{(0)}|^2$, where $u^{(0)}$ is the solution to eq. \eqref{NLS_st}$\big|_{\la=0}$.}  
only when $\la\sim\nu L$. 
Accordingly, we choose 
\be\lbl{la_sc}
\la = \eps\nu L\,,
\ee 
where $\eps$ is a sufficiently small but (for the moment) {\it fixed} constant, responsible for the effective size of nonlinearity, in the sense that it governs the size of nonlinearity in the WKE (see below). 
The parameter $\eps$ should be small for the methodology of the WT to apply.

We introduce the damped/driven (modified) WKE as
\be\label{WKE}
\dot \fm(s,\tau) = \eps^2 K(s,\tau)\big(\fm(\cdot,\tau)\big) -2\ga_s\fm(s,\tau) + 2 b(s)^2,  \qquad  \;\; \fm(s,0)=0,
\ee
where $s\in\R^d$ and $\tau\ge 0$, while $K(s,\tau)$ is a non-autonomous cubic nonlinear integral operator, called the {\it wave kinetic integral} and defined in Appendix~\ref{a:WKI}.
In \cite{DKMV} it is shown that for small $\eps$ the WKE \eqref{WKE}  admits a unique solution, and the latter has the form 
\be\lbl{zz}
\fm(s,\tau) = \fm^{(0)}(s,\tau) + \eps^2\fm^{(1)}(s,\tau;\eps), \qmb{where}\qu \fm^{(0)}, \fm^{(1)}\sim 1
\ee
 and $\fm^{(0)}(s,\tau)$ solves the linear equation \eqref{WKE}$\big|_{\eps=0}$.

Let us decompose the solution to eq. \eqref{NLS_st}, \eqref{in_data}, \eqref{la_sc} as
\be\lbl{decomp_u}
u(x,\tau)  = U^{M}(x,\tau) + Err^M(x,\tau), \qmb{where}\qu  U^M = \sum_{k=0}^M  \eps^k u^{(k)}(x,\tau),\qu M\ge 0,
\ee
$u^{(k)}$ are terms of the formal decomposition of the solution $u$ in $\eps$ and $Err^M$ is a reminder.  So, the Gaussian process $u^{(0)}$ is a solution to the linear equation \eqref{NLS_st}$\big|_{\la=0}$, while the higher order terms $u^{(k)}$, $k\ge 1$, are computed iteratively; they are Wiener chaoses of orders $2k+1$.
We call the process $U^M$ the {\it quasisolution} of order $M$.
\footnote{By analogy with the quasimodes in the spectral theory of the Shr\"odinger operator.}

 Together with the energy spectrum $\frak n_s(\tau)$ of the exact solution $u(\tau)$, given by \eqref{es},  let us consider that of the quasisolution
 \[
 \frak N^{M}_s(\tau) = \EE|\hat U^{M}_s(\tau)|^2,\qquad \hat U^{M} = \cF(U^{M}).
 \]
 The first limit in \eqref{lim}, when $\nu\to 0$ and $L$ stays fixed, is known as the {\it limit of discrete turbulence} \cite[Section  10]{Naz11} and was successfully studied in \cite{HKM, KM16}. 
 It was shown there that under this limit distributions of solutions to \eqref{NLS_st} are governed in appropriate sense by those to the {\it effective equation of  discrete turbulence} \eqref{eff}, see for details Section~\ref{s:dt}.
 In particular, this implies the convergence 
 \[
 \frak n_s (\nu,L;\tau) \to n_s(L;\tau),\qquad \frak N^M_s(\nu,L;\tau) \to N^M_s(L;\tau)\qu \forall M\ge 0  \qmb{as}\qu \nu\to 0
 \]
 on finite time intervals,
 where $n_s(L;\tau)$ and $N_s^M(L;\tau)$ are the energy spectrums of the solution and quasisolution to eq. \eqref{eff}.
  Thus, it remains to study their limits as $L\to\infty$.

Below we denote by $C^\#(s)$ various positive functions decaying when $|s|\to \infty$ faster than $|s|^{-r}$ for any $r>0$, and by $C_{p}^\#(s)$ the functions $C^\#(s)$ depending on a parameter $p$.
Let $\fm$ be the unique solution to eq. \eqref{WKE}.
 The main result of \cite{DKMV} is: 
 
 \msk
\noindent {\bf Theorem A.} (\cite{DKMV}). {\it The energy spectrum $N_s^{2}$
	 satisfies
	$$
	|N^2_s(L;\tau) - \fm(s,\tau)| \le  C^\#(s)\eps^3,\qquad s\in\Z^d_L,
	$$
	uniformly in $\tau\ge 0$, for sufficiently small $\eps$ and large $L$}.
\msk

See Theorem~\ref{t:DKMV} below for details. 
One of the key steps in proving Theorem~A was obtaining uniform in $L$ estimates for the Fourier transforms of the terms $\hat u^{(k)} = \cF(u^{(k)})$  of decomposition \eqref{decomp_u}.
\footnote{More precisely, for the terms $a_s^{(k)}$  of decomposition \eqref{a-A-w} for a solution to the effective equation \eqref{eff}.} 
In \cite{DKMV}, using the Feynman diagrams and two powerful results from the number theory --- the Heath-Brown circle method and the finite fields' Bezout theorem --- we were able to get the desired estimates only for $k\le 2$ (which was sufficient for studying the energy spectrum  $N_s^2$ of  the 2-nd order quasisolution).
In the present paper we get the estimates for $\hat u_s^{(k)}$ with any $k\ge 0$, employing completely different and much simpler argument, based on inductive  analysis of cumulants of the random variables $\hat u_s^{(k)}$. We still need the  circle method but do not use neither the Bezout theorem, nor the Feynman diagrams any more. See Proposition~\ref{l:bound_a} and Section~\ref{s:Fd-cum} for a discussion. 
The obtained estimates immediately lead to 

\msk
\noindent{\bf Proposition B.} {\it Assertion of Theorem A holds for the energy spectrum $N_s^{M}$ with arbitrary $M \ge 2$.}
\footnote{With the same precision $\eps^3$ of the approximation.}
\msk

See Proposition~\ref{t:DKMV-extention}. Now we are aiming to extend this result to the energy spectrum $n_s(L;\tau)$ of the exact solution to the effective equation \eqref{eff}. To this end we choose
\be\lbl{subcr}
\eps = L^{-\al}, \qquad 0<\al\le\frac12.
\ee
Following paper \cite{DH} (in which a related problem was studied, see Section~\ref{s:res_dissc}), we call this scaling {\it subcritical}, while the  scaling under which $\eps$ is a fixed small constant, independent from  $L$, is called {\it critical}.
Work under the subcritical scaling is technically much simpler than under the critical one, but is still highly non-trivial. 

\msk

\noindent{\bf Main Theorem.} {\it Assume that $d\ge 3$ and $r_*>d-1$. Then for any $T> 0$, under  subcritical scaling \eqref{subcr}   the energy spectrum $n_s(L;\tau)$  satisfies 
	$$
	|n_s(L;\tau) - \fm(s,\tau)| \le  C_{\al,T}^\#(s)\eps^3, \qquad s\in\Z^d_L,
	$$
	uniformly in $\tau \in [0, T]$, for sufficiently large $L$}.
\msk

See Theorem~\ref{t:main} for details. To get this result we show that quasisolutions of high order to the effective equation \eqref{eff}  well approximate the exact solutions and then employ Proposition~B. The proof of the former assertion relies on the discussed above uniform in $L$ estimates and similar ones for linear operators, obtained by linearizing the nonlinearity in eq. \eqref{eff} at the quasisolutions.

The restriction $r_*>d-1$ arises primarily from the number-theoretic bound in Corollary~\ref{l:corr_nt}. We expect that this bound can be improved, see Conjecture~\ref{conj:nt}.  If so, it should be possible to extend our results to all~$r_*>0$.

Our paper is self-contained modulo Theorem A and a number theoretic result from Appendix~\ref{a:nt}, which we borrow from \cite{DKMV} and \cite{number_theory} correspondingly.
Note that we do not use Theorem A in its full strength, see Remark~\ref{r:useDKMV} below.

\subsection{Brief literature review}
\lbl{s:res_dissc}

There are plenty of physical works developing some different (but consistent) approaches to the study of the WT kinetic limit \eqref{limmm} for eq. \eqref{NLS}, mostly with $\nu=0$, see \cite{ZLF92, Naz11, NR} and references therein. 
Despite long-standing significant interest in the community of mathematical physicists to the problem of their rigorous justification, mathematical works devoted to this question started to appear only a decade ago \cite{BGHS18, BGHS, CG1,CG2,DH,DH1,DH1as,DH1lt,DK1,DK2,DKMV,Faou,HG, HRST, LS, ST, AS}.  E.g., see introductions to \cite{DH1, DK1, HG} for rather detailed state of the art  reviews.

Probably the most impressive success was achieved in a series of papers \cite{BGHS, CG1,CG2,DH,DH1,DH1as,DH1lt} studying NLS equations, similar to eq. \eqref{NLS}$\big|_{\nu=0}$ with random initial data, in which limit \eqref{limmm} was specified as $L\to\infty$ and $\la\to 0$ simultaneously under various appropriate scalings. 
More specifically, in \cite{BGHS}, under a {\it subcritical} scaling of parameters, the authors proved that the energy spectrum of an exact solution is well approximated by a solution to the (non-modified) WKE. In \cite{CG1,CG2, DH}  this result was improved to less restrictive but still {\it subcritical} scalings similar to \eqref{subcr}. 
In \cite{DH1,DH1as} Deng and Hani extended these results to the {\it critical} scaling, and finally in \cite{DH1lt} they managed to get rid of a small time assumption (while in the previous works it was assumed that the considered slow-time interval is small). 

Rigorous investigation of eq. \eqref{NLS} in the Zakharov-L'vov setting, i.e. with $\nu\ne 0$, was started in \cite{DK1, DK2} (see also review \cite{DKsmall}). There the limit \eqref{lim} was studied  in the opposite to \eqref{lim} regime, when $L\to\infty$ much faster than $\la^{-1}$, $\nu^{-1}$ do. It was shown that the energy spectrums $\frak N^M$ of the quasisolutions $U^M$ of orders $2\le M\le d$ are well approximated by a solution to the (non-modified) WKE. As we have already discussed, in \cite{DKMV} a similar result was obtained  under limit \eqref{lim} for $M=2$ with the  modified WKE \eqref{WKE}.
In a recent paper \cite{HG}, using the technique developed in \cite{DH}, the authors studied eq. \eqref{NLS} in a regime similar to that from \cite{DK1}. They proved the desired kinetic approximation for the exact solution under a \textit{subcritical} scaling. 
\footnote{Another difference of \cite{HG} and \cite{DK1} is that Grande and Hani consider random initial conditions instead of zero ones as in \cite{DK1}.  They also study a regime in which the noise and damping are so small that the kinetic description coincides with that from \cite{DH} where the deterministic NLS with random initial data was studied, as well as a regime when the stochastic perturbation dominates, so that the limiting dynamics does not undergo nonlinear effects.}
The authors claimed that the result can be extended to the \textit{critical} scaling using the technique developed in \cite{DH1}, but the proof was not given.

\subsection{Methods}
\lbl{s:Fd-cum}
The principle ingredient of all works cited above is estimating of the terms of decomposition~\eqref{decomp_u}, which, due to our knowledge, is always based on a method of Feynman diagrams going back to \cite{EY00,ESY07}. 
The latter suggests, by iterating the Duhamel formula, to express the processes $\hat u_s^{(k)} = \cF(u^{(k)})_s$ via the Gaussian processes $\hat u_s^{(0)}$, to parametrize the resulting sums, giving $\hat u_s^{(k)}$, by trees and then to compute the correlations $\EE \hat u_s^{(k)} \bar {\hat u}_s^{(n)}$, forming the energy spectrum, via the Wick theorem. 
This leads to summations, naturally parametrized by trees,
whose height grows with $k$ and $n$, 
coupled via their leaves, usually called the Feynman diagrams. 

Once such diagrammatic presentation is developed, one estimates components of the correlation $C_{kn}(s):=\EE \hat u_s^{(k)}\bar {\hat u}_s^{(n)}$, corresponding to each diagram,  and finds strong cancellations between the components, corresponding to different diagrams. 
Realization of this program is complicated due to the growth with $k+n$  of size of the Feynman diagrams and the fast growth of the number of different diagrams.
It turns out to be technically very involved already at the level of quasisolutions \cite{DK1,DK2,DKMV} or under the subcritical scalings \cite{BGHS, CG1,CG2,DH,HG}. Under the critical scaling this scheme leads to extremely sophisticated and long proofs \cite{DH1,DH1as,DH1lt}.

In \cite{DKMV} this strategy allowed to represent the limit as $\nu\to 0$ of the correlations
$C_{kn}(s)$
 in the form
\be\lbl{Feynm}
\sum_{\cF} L^{N(1-d)}\sum_{z\in \Om^\cF \cap \Z^d_L} \Phi^\cF(s,z),\qquad N:=k+n,
\ee
where the functions $ \Phi^\cF$ are Schwartz.
The first summation here is taken over an appropriate (finite) subset of Feynman diagrams while $\Om^\cF= \cap_{j=1}^{N-1} \om^\cF_j$, where $\om_j^\cF$, $1\le j\le N-1$, are certain explicitly written non-degenerate linearly independent quadrics. In the case $N\le 4$ in \cite{DKMV} the sums from \eqref{Feynm} were analysed via the circle method of Heath-Brown \cite{HB,number_theory} joined with the Bezout theorem for finite fields. For large $N$, however, we were not able to follow this strategy since for that one needs to prove that the quadrics $\om_j^\cF$ intersect transversally, which seems to be a highly non-trivial algebraic geometry problem, see \cite{Vlad} and \cite[Section 3.2]{DKMV}.

In the present paper we \textit{do not} use the Feynman diagrams.  
 Instead we inductively bound cumulants of the Fourier coefficients $\hat u^{(k)}_s$,
 \footnote{More precisely, those of terms $a^{(k)}_s$ of decomposition \eqref{a-A-w}.}
 which requires analysing 
 sums as in \eqref{Feynm} only in the simple case $N=2$. 
 An advantage of using cumulants over moments is that they effectively separate contributions of different scales in $L$, see Remark~\ref{r:cum-mom}. 
 Our analysis is partially inspired by paper \cite{LM}, where the authors have written down a hierarchy of equations for the time-evolution of cumulants under a discrete NLS equation (although we do not use the obtained there results).

 It is very likely that the suggested relatively simple approach applies to other orders of limits in \eqref{limmm}  as well as for analysis of eq.~\eqref{NLS} with $\nu=0$ and random initial data. We hope as well that a refinement of our proof could allow to establish the WT kinetic approximation under the {\it critical} scaling, see Remark \ref{r:constants}(ii). 


\subsection{Notation}
\lbl{s:not}

{\it (1)} By $C,C',C_1,\dots$ we denote various positive constants. 
By $C^\#(x), C_1^\#(x), \wt C^\#(x), \dots$ we denote various non-negative fast decaying functions on $\R^m$, $m\ge 1$, satisfying for every $n\ge 1$
\[
C^\#(x) \le C_n \lan x\ran^{-n} \qquad \forall x\in \R^m, 
\]
where 
\be\non
\lan x\ran:=\max(1, |x|).
\ee
 Constants $C,C_1,\dots$ and functions $C^\#,C_1^\#,\dots$ may vary from formula to formula. 
By $C_p,\dots$ and $C_p^\#(x),\dots$ we denote the constants $C$ and functions $C^\#$ depending on a parameter $p$. Usually we do not indicate the dependence on the dimension $d\ge 3$, time  $T>0$, viscosity constant $r_*>0$ and the scaling parameter $\al>0$ from \eqref{subcr}.

{\it Properties of functions $C^\#$} (see \cite[Section 1.4]{DK1}). 

$\bullet$ If $Q: \R^m\mapsto\R^m$ is a linear isomorphism, the function $C^\#(Qx)$ can be written as $C^\#_1(x)$. 

$\bullet$ For any function $C^\#(x,y)$ with $(x,y)\in\R^{m_1+m_2}$ there exist functions $C_1^\#(x)$ and $C^\#_2(y)$ such that $C^\#(x,y) \le C_1^\#(x) C_2^\#(y)$. 
Vice versa, for any functions $C_1^\#(x)$ and $C^\#_2(y)$ the function $ C_1^\#(x) C_2^\#(y)$ can be written as $C^\#(x,y)$.
   
{\it (2)}
For a set $X$ we denote by $|X|$ its number of elements. If $X$ is a subset of a vector space, we write $\sum X:= \sum_{x\in X} x$.

{\it (3)}
Let $X$ and $Y$ be disjoint sets. By $\cP(X)$ we denote the set of all partitions $\pi$ of $X$. By $\cP_2(X\sqcup Y)$~---  that of all partitions $\pi\in\cP(X\sqcup Y)$ such that $A\cap Y \ne\emptyset$  for any $A\in\pi$. Note that in general $\cP_2(X\sqcup Y) \ne \cP_2(Y\sqcup X)$.

{\it (4)} For a vector $\bm v=(v_1,\dots,v_n)$ we denote $\bm v_{\ge j}:=(v_j, v_{j+1,}\dots, v_n)$.
\msk 

{\bf Acknowledgments.} I am deeply grateful to Sergei B. Kuksin and Alberto M. Maiocchi for our many discussions of the problem.

\section{Results in detail and schemes of their proofs}
\lbl{s:res_det}

\subsection{The limit of discrete turbulence $\nu\to 0$}
\lbl{s:dt}

Let us  assume scaling \eqref{la_sc}, where for the moment $\eps>0$ is a fixed constant.
We start by discussing the first limit, called the  {\it limit  of discrete turbulence},
\be\lbl{lim_1}
\nu\to 0 \qnd  \mbox{$L$ stays fixed},
\ee
studied in  \cite{HKM,KM16}. Let us substitute the Fourier series \eqref{Fourier-def} to eq. \eqref{NLS_st} and  pass to the {\it interaction representation}
\be\lbl{interaction r}
a_s(\tau)=\hat u_s(\tau) e^{-i\nu^{-1}\tau |s|^2}, \qu s\in\Z^d_L,
\ee
that kills the linear term $\F(\nu^{-1}i\Del u)$ in eq. \eqref{NLS_st}, written in the Fourier presentation. Then eq. \eqref{NLS_st}, \eqref{diss_op}, \eqref{in_data} with $\la$ given by \eqref{la_sc} takes the form
\begin{equation}\label{a-eq}
	\begin{split}
		\dot a_s+\gamma_s a_s &= i\eps{\mathbb Y}_s(a, \nu^{-1}\tau)
		+b(s) \dot\beta_s, \qquad a_s(0) = 0
		\,,\qquad s\in{{\mathbb Z}}^d_L  \,,\\
		{\mathbb Y}_s(a,t)&=L^{-d+1} \sum_{s_1,s_2,s_3\in\Z^d_L} \de'^{s_1s_2}_{s_3s}\, \exp\big(it\omega^{s_1s_2}_{s_3s}\big)\, 
		a_{s_1} a_{s_2} \bar a_{s_3}\,,
	\end{split}
\end{equation}
where $\{\beta_s(\tau)\}$ is another set of standard complex independent Brownian motions, the bar denotes the complex conjugation,
\be\lbl{def_del}
{\delta'}^{s_1s_2}_{s_3s}: =
\left\{\begin{array}{ll}
	1,& \text{ if $s_1+s_2=s_3+s$ and $s_1, s_2 \ne s_3$}
	\,,
	\\
	-1,& \text{ if $s_1=s_2=s_3=s$}\, , \\ 
	0,  & \text{otherwise}
\end{array}\right.
\ee
and
\be\lbl{oms}
\om^{s_1s_2}_{s_3s}:=  |s_1|^2 + |s_2|^2 - |s_3|^2 - |s|^2 = -2(s_1-s)\cdot(s_2-s)\, .
\ee
Here in the last equality we expressed $s_3$ from \eqref{def_del} as $s_3=s_1+s_2-s$.

Since $|a_s| = |\hat u_s|$, the energy spectrum \eqref{es} 
takes the form
\be\non
\frak n_s(\tau) = \EE |a_s(\tau)|^2, \qu s\in\Z^d_L.
\ee
Together with eq. \eqref{a-eq} we consider the {\it effective equation of discrete turbulence}. The latter is obtained from \eqref{a-eq} by averaging the nonlinearity $\mathbb Y(a,t)$ in $t$, that is by keeping only resonant terms in the sum from \eqref{a-eq}:
\begin{equation}\label{eff}
	\begin{split}
		\dot a_s+\gamma_s a_s &= i\eps Y_s(a)
		+b(s) \dot\beta_s, \qquad a_s(0) = 0
		\,,\qquad s\in{{\mathbb Z}}^d_L  \,,\\
		Y_s(a)&=L^{-d+1} \sum_{s_1,s_2,s_3\in\Z^d_L}\Del^{s_1s_2}_{s_3s}
		a_{s_1} a_{s_2} \bar a_{s_3}\,, \qquad \Del^{s_1s_2}_{s_3s}:= \de'^{s_1s_2}_{s_3s} \de(\omega^{s_1s_2}_{s_3s}),
	\end{split}
\end{equation}
where the Kronecker delta $\de(x)= 1$ if $x=0$ and  $\de(x)= 0$ otherwise.
 
In \cite{HKM,KM16} it is shown that, once $r_*$ from \eqref{gamma} is sufficiently large,  equations  \eqref{a-eq}  and \eqref{eff} have unique solutions $a^\nu(\tau)$ and $a(\tau)$, and under limit \eqref{lim_1} the solution $a^\nu(\tau)$ converges in distribution to  $a(\tau)$ on finite time intervals. See Theorem 1.1 from \cite{DKMV} and a subsequent discussion. 
Moreover, it is shown that for any  $m\ge 0$
\be\lbl{apr_bound}
\EE \sup_{\tau\in[0,T]}\big|a_s^\nu(\tau)|^m , \qu \EE\sup_{\tau\in[0,T]} \big|a_s(\tau)|^m \;\le \; L^k C_{m}^\#(s), \qquad  \forall s\in\Z^d_L,
\ee
for some $k=k(m)\ge 0$ uniformly in $\nu$.
More precisely, in \cite{HKM, KM16} the well-posedness and bound \eqref{apr_bound} were proved only when  $\frak U(u) = (1-\Del)u$ and $d\le 3$ but it was argued that they can be similarly established for arbitrary $d$ if the operator $\frak U$ provides sufficiently strong damping, i.e. $r_*\ge r_*(d)$ is sufficiently large. The dependence of estimate \eqref{apr_bound}  on $L$ was not explicitly written but it can be obtained by a straightforward analysis of the proof. 
\footnote{In our main result, Theorem~\ref{t:main}, we assume some of these results only in item (2).}

These results immediately imply that under limit \eqref{lim_1} the energy spectrum $\frak n_s(\tau) = \frak n_s(\nu;\tau)$ of the solution $a^\nu(\tau)$ to eq. \eqref{a-eq} converges to that of the solution $a(\tau)$ to eq. \eqref{eff},
\[ 
 \EE |a^\nu_s(\tau)|^2 \to \EE |a_s(\tau)|^2 \qmb{as}\qu \nu\to 0, \qquad s\in \Z^d_L,
\]
uniformly on finite time intervals.

\subsection{The main result}
In the present paper we study the limiting as $L\to\infty$ behaviour of the energy spectrum 
	$
	\EE |a_s(\tau)|^2 = 	\EE |a_s(L;\tau)|^2
	$
of the solution $a_s(\tau)$ to the effective equation of discrete turbulence \eqref{eff}.	

Let us denote by $\frak X_r$, $r\ge 0$, a   space of vector-functions $v=\{v_s(\tau),\; \tau\ge 0, \, s\in\Z^d_L\}$ with continuous components $v_s\in C([0,T],\C)$, provided with the normalized weighted $l^1$-norm
\be\lbl{r-norm}
|v|_{\frak X_r}:=L^{-d}\sum_{s\in\Z^d_L}\lan s\ran^r|v_s|_\infty  \qmb{where}\qu |v_s|_\infty:= \sup\limits_{\tau\in[0,T]}|v_s(\tau)|\, .
\ee
Our main result is the following theorem.
For an event $Q$ and a random variable $\xi$ we denote $\EE_Q\xi:=\EE \big(\mI_{Q}\xi\big)$, where $\mI_{Q}$ stands for the indicator function of  $Q$.

	\begin{theorem}\lbl{t:main}
			Let $d\ge 3$. Assume 
		subcritical scaling \eqref{subcr} with any  $0<\al\le1/2$ and that $r_*>d-1$. Then 
		
	(i)	For any  $r\ge 0$ and $L\ge L_{r}$ with sufficiently large $L_r$ there is an event $\Om_{L}=\Om_L(r)$ satisfying
		$\PP (\Om_{L}) \ge 1- C_r^\#(L)$, 
		in which eq. \eqref{eff} has a unique  solution $a\in\frak X_r$. The solution satisfies
		\be\lbl{b_main}
	\sup\limits_{\tau\in [0,T]}	\Big|\EE_{\Om_{L}} |a_s(\tau)|^2 - \fm(s,\tau) \Big| \le C_{r}\lan s \ran^{-r} \eps^3, \qquad\qu \forall s\in\Z^d_L, 
		\ee
		 where  $\fm(s,\tau)$ is a unique solution to  WKE \eqref{WKE}.
		
	(ii) Assume that eq. \eqref{eff} has a solution
	\footnote{Defined not only in the event $\Om_L$ but everywhere.}
	 $a_s(\tau)$, satisfying estimate \eqref{apr_bound} with some $m>2$ and $k\ge 0$. Then approximation \eqref{b_main} holds with the expectation $\EE_{\Om_L}$ replaced by $\EE$, i.e. for the energy spectrum $n_s(L;\tau) = \EE |a_s(\tau)|^2$.
	\end{theorem}	
	Next we give a detailed scheme of the proof of Theorem~\ref{t:main}.
	
\subsection{Analysis of quasisolutions to eq. \eqref{eff}}

For  complex vectors $v^j = (v^j_s)_{s\in\Z^d_L}$ let us set
\be\lbl{Ydef}
Y_s(v^1, v^2, v^3)= 
L^{-d+1}  \sum_{s_1,s_2,s_3\in\Z^d_L}\Del^{s_1s_2}_{s_3s} 
\, v^1_{s_1}v^2_{s_2}\bar v^3_{s_3},
\qquad s\in\Z^d_L,
\ee 
so that $Y_s(v) = Y_s(v,v,v)$ (see \eqref{eff}).
If $v_s^j=v_s^j(\tau)$ are continuous functions of time, we set 
\be\lbl{cYdef}
\cY_s(v^1, v^2, v^3)(\tau) = 
\int_0^\tau e^{-\ga_s(\tau-l)}\, Y_s(v^1,v^2,v^3)(l)\, dl,
\qquad s\in\Z^d_L,
\ee 
and $\cY= \{\cY_s, \,s\in\Z^d_L\}$.
We also denote
\be\lbl{Y^sym}
Y^{sym}_s(a,b,c) := Y_s(a,b,c) + Y_s(c,a,b) + Y_s(b,c,a)
\ee
and similarly define $\cY^{sym}_s$. 

\ssk
 	
For $M\ge 0$ let us write  the solution $a_s(\tau)$ of the effective equation \eqref{eff} in the form
\be\lbl{a-A-w}
a_s(\tau) = A_s^M + w_s^M, \qmb{where}\qu A_s^M := \sum_{m=0}^M \eps^m a_s^{(m)}(\tau), 
\ee
the processes $a_s^{(m)}(\tau)=a_s^{(m)}(L;\tau)$ are independent from $\eps$ and $w_s^M(\eps,L;\tau)$ are reminders. 
The processes $a_s^{(0)}(\tau)$
solve the linear equations
$$
\dot a_s^{(0)} + \gamma_s a_s^{(0)}
= b(s) \dot\beta_s  \,,\quad s\in{{\mathbb Z}}^d_L\,,
$$
and thus have the form
\be\label{a0}
a^{(0)}_s(\tau) = b(s) \int_{0}^\tau e^{-\gamma_s(\tau-l)}d\beta_s(l), \qquad s\in\Z^d_L.
\ee
These are a Gaussian processes, independent for different $s$. 
By a straightforward computation (see (2.8) in \cite{DK1}) we find that 
for any $s,k\in \Z^d_L$ and $\tau_1,\tau_2\ge 0$
\be\lbl{a^0-corr} 
\begin{split}
	& \EE a_s^{(0)}(\tau_1)  a_{k}^{(0)}(\tau_2) =  \EE \bar a_s^{(0)}(\tau_1) \bar a_{k}^{(0)}(\tau_2)= 0,
	\\ 
	& \EE a_s^{(0)}(\tau_1) \bar a_{k}^{(0)}(\tau_2)
	= \delta^s_{k} \,  \frac{b(s)^2}{\gamma_s} \big( e^{-\gamma_s|\tau_1-\tau_2|} -
	e^{-\ga_s(\tau_1+ \tau_2)}\big),
\end{split}
\ee
where $\de_s^s=1$ and $\de_s^k=0$ for $s\ne k$.
The processes $a_s^{(m)}$ with $m\ge 1$ are obtained iteratively,
\be\label{an}
\begin{split}
	a^{(m)}_s&(\tau) 
	= i\sum_{m_1+m_2+m_3=m-1}   \cY_s(a^{(m_1)},a^{(m_2)}, a^{(m_3)} )(\tau), \qquad s\in\Z^d_L,
\end{split}
\ee
and are Wiener chaoses of orders $2m+1$.

We call the process $A^{M}$ {\it the  quasisolution of order $M$} to eq. \eqref{eff}.
The following theorem, in which we view $\eps$ as an {\it independent} from $L$ small parameter, is the
main result of \cite{DKMV}. 
Recall that $\fm(s,\tau)$ denotes the unique solution to the WKE \eqref{WKE}.
\begin{theorem}[\cite{DKMV}]\lbl{t:DKMV} Let $d\ge 3$.
Then for any $r\ge 0$ and $M=2$,
\be\lbl{DKMV}
\Big| \EE|A^{M}_s(\tau)|^2 - \fm(s,\tau) \Big|\le  C_{r}\lan s \ran^{-r} \eps^3, \qquad \forall s\in\Z^d_L,
\ee
uniformly in $\tau \ge 0$, $\eps\le \eps_r$ and $L\ge \eps^{-2}$, with sufficiently small $\eps_r>0$.
\end{theorem}

 Theorem \ref{t:DKMV}  was proved in \cite{DKMV} only for $M=2$ since  we were not able to obtain uniform in  $L$ estimates for the terms $a_s^{(m)}$ with large $m$.
 Now we are able to get them: the key ingredient of  the present work is the  following bound. 
\begin{proposition} \lbl{l:bound_a} 
	For any $m\ge 0$ and $q\ge 1$  
	\begin{align}\lbl{bound_a}
		\EE \sup_{\tau\le l \le \tau+1}\big|a_s^{(m)}(l)\big|^{2q} \leq C_{m,q}^\#(s) \, ,
	\end{align}
uniformly in $\tau\ge 0$ and $L\ge 1$.
\end{proposition}
Proposition~\ref{l:bound_a} is deduced in Section~\ref{s:est_a} as a corollary of a stronger result, Proposition~\ref{t:sup_est_num}, in which we bound joint cumulants of the random variables $a_s^{(m)}$. In Appendix~\ref{a:const_growth} we show that the  function $C_{m,q}^\#(s)$ has at most factorial growth in $m$ and $q$ (we do not need this to establish Theorem~\ref{t:main}), see Remark~\ref{r:constants} below. Proof of this fact  requires an additional combinatorial argument.
  
Due to estimate \eqref{bound_a}, for any $M\ge 2$
\[
\big| \EE|A_s^M(\tau)|^2 - \EE|A_s^2(\tau)|^2 \big| \le C_M^\#(s)   \eps^3,
\]
uniformly in $\tau\ge 0$, $L\ge 1$ and $\eps\le 1$.
This bound  immediately  extends the result of \cite{DKMV} to the quasisolutions $A^M$ of arbitrary order:
\begin{proposition} \lbl{t:DKMV-extention}
	Assertion of Theorem~\ref{t:DKMV} holds for any $M\ge 2$ (with the constants $C_{r}$ depending on $M$).
\end{proposition}

\begin{remark}\lbl{r:useDKMV}
	To get Proposition~\ref{t:DKMV-extention} we do not use Theorem~\ref{t:DKMV} in its full strength. Indeed, in \cite{DKMV} the energy spectrum $ \EE|A_s^2(\tau)|^2$  of the 2-nd order quasisolution $A^2$ to eq.~\eqref{eff} was written in the form 
	$N_s^{\le 2} + N_s^{>2}$, where $N_s^{\le 2} = \sum_{m+n\le 2}\eps^{m+n}\EE a_s^{(m)}\bar a_s^{(n)}$ and 
	$N_s^{> 2}$ is given by a similar sum over $\{m+n=3,4: m,n\le 2\}$.
	Using the circle method, in \cite[Sections 4,5]{DKMV} it was proven that the term $N^{\le 2}$ leads to the wave kinetic equation --- this fact is needed to get Proposition~\ref{t:DKMV-extention}. 
	In \cite[Sections 2,3]{DKMV}, using the Feynman diagrams, the circle method and the Bezout theorem for finite fields, 
	it was shown that $N_s^{> 2} \le \eps^3 C^\#(s)$, so the term $N_s^{> 2}$ can  be viewed as a reminder in \eqref{DKMV}. We do not need this bound any more since it immediately follows from Proposition~\ref{l:bound_a}. 
\end{remark}

\subsubsection*{\underline{Sketch of the proof of Proposition~\ref{l:bound_a}}}

Let us consider a multi-set 
\be\lbl{J_intro}
J=\Big\{a_{\xi_1}^{(m_1)}(\tau_1), \dots, a_{\xi_p}^{(m_p)}(\tau_p), \bar a_{\s_1}^{(n_1)}(\tau'_1), \dots,\bar a_{\s_{p'}}^{(n_{p'})}(\tau'_{p'})\Big\}
\ee
with $p,p'\ge 0,$  $m_i,n_j\ge 0$, $\xi_i,\s_j\in\Z^d_L$ and $\tau_i,\tau'_j\ge 0$.  We will study the cumulant $\ka(J)$, 
see Appendix~\ref{a:cum} for a reminder about the cumulants.
It is not difficult to see that
\be\lbl{base_prop_intro}
\ka(J)\ne 0 \quad \Rightarrow \quad (i) \; p'=p \qnd (ii)\; \sum_{j=1}^p \xi_j =  \sum_{j=1}^p \s_j, 
\ee
so further on we assume $p'=p$. 
Setting $N:=\sum_{j=1}^p (m_j+n_j)$, we claim that 
\be\lbl{ind_as_intro}
|\ka(J)| \le C_{p,N}^\#(\bxi,\bsi) L^{-(d-1)(p-1)}, 
\ee
where $\bxi = (\xi_j)_{1\le j\le p}$ and $\bsi = (\s_j)_{1\le j\le p}$. In particular \eqref{ind_as_intro} with $p=1$ implies 
$\EE|a_s^{(m)}|^2 = \ka\big(\{a_s^{(m)}, \bar a_s^{(m)}\}\big) \le C_m^\#(s)$,
while bounds for the expectations $\EE|a_s^{(m)}|^{2q}$ follow from \eqref{ind_as_intro} by the Leonov-Shiryaev formula \eqref{mom-cum}. Estimate \eqref{bound_a} is then deduced from the latter bounds and similar ones in a standard way.

To prove \eqref{ind_as_intro}, we argue   by induction in $N$.
The base is trivial since for $m_j=n_j=0$ the set $J$ consists of Gaussian processes. 
Assume now for definiteness that $m_1\ge 1$. 
Let $\check J:= J\sm\{a_{\xi_1}^{(m_1)}\}$. Due to \eqref{an} and multi-linearity of cumulants, 
\[
|\ka(J)| \le\sum_{k_1 + k_2 + k_3=m_1-1} \sup_{0\le l\le \tau_1}|K(l)|,
\]
where 
\[
K(l) = L^{-d+1} \sum_{s_1,s_2,s_3\in\Z^d_L} \Del^{s_1s_2}_{s_3s}\, \ka\Big(\check J\sqcup \Big\{\big(a_{s_1}^{(k_1)}a_{s_2}^{(k_2)}\bar a_{s_3}^{(k_3)}\big)(l)\Big\}\Big).
\]
Due to the Malyshev's formula \eqref{Mal}, 
\be\lbl{Mal_intro}
\ka\Big(\check J\sqcup \Big\{\big(a_{s_1}^{(k_1)}a_{s_2}^{(k_2)}\bar a_{s_3}^{(k_3)}\big)(l)\Big\}\Big)= 
\sum_{\pi\in\cP_2(\check J\sqcup J_{new})} \prod_{\cA\in\pi} \ka(\cA),
\ee
where $J_{new}:=\Big\{a_{s_1}^{(k_1)}(l), a_{s_2}^{(k_2)}(l),\bar a_{s_3}^{(k_3)}(l)\Big\}$ and the set of partitions $\cP_2$ is defined in Notation~\ref{s:not}(3).
Since for any $\pi\in\cP_2(\check J\sqcup J_{new})$ and $\cA\in\pi$ we have $\cA\cap J_{new}\ne\emptyset$, the partition $\pi$ consists of at most three elements, $1\le|\pi|\le 3$. 
Assume first that $|\pi|=1$, so that $\pi=\{\check J\sqcup J_{new}\}$. Since the multi-set $\check J\sqcup J_{new}$ has the form \eqref{J_intro} with $p:=p+1$, by the induction assumption $|\ka(\check J\sqcup J_{new})| \le C^\#(s_1,s_2,s_3,\bxi_{\ge 2},\bsi) L^{-(d-1)p}$ (see Notation~\ref{s:not}(4)).   Then, 
\[
\begin{split}
|K(l)| &\le L^{-d+1} \sum_{s_1,s_2,s_3} \Del^{s_1s_2}_{s_3s}\, C^\#(s_1,s_2,s_3,\bxi_{\ge 2},\bsi)L^{-(d-1)p} \\
 &\le C^\#(\bxi,\bsi) L^{-(d-1)(p - 1)},
 \end{split}
\]
Here we used the bound of Corollary~\ref{l:corr_nt} (with $a=b=0$),  following from the circle method, for the summation over the quadric $\{\om^{s_1s_2}_{s_3s}=-2u \cdot v =0\}$ (see \eqref{oms}), where $u:=s_1-s$, $v:=s_2-s$. This 
agrees with \eqref{ind_as_intro}.
 
The cases $|\pi|=2,3$ are considered similarly, additionally using   that the product from the r.h.s. of  \eqref{Mal_intro} vanishes unless the identity from item \textit{(ii)} of \eqref{base_prop_intro} holds for the every set $\cA\in\pi$.

\begin{remark}\lbl{r:cum-mom} 
	The inductive procedure in the sketch above is effectively run on cumulants rather than on moments since
	the cumulants $\ka(J)$  are at most of size $O_{\bxi,\bsi}(L^{-(d-1)(p-1)})$ for all vectors of indices $\bxi,\bsi$, see \eqref{ind_as_intro}.\footnote{In fact the scale $L^{-(d-1)(p-1)}$ is achieved only for some vectors of indices $\bxi,\bsi$ while for the other the cumulants $\ka(J)$ should be smaller, see in the end of Remark~\ref{r:constants}(ii). We do not need this refinement in the present work.}
	This bound suffices for our goal. 
	 On the contrary, the moments $\EE \prod_{u\in J}u$ are of size $O_{\bxi,\bsi}(L^{-k_{d,p}(\bxi,\bsi)})$ with $k_{d,p}(\bxi,\bsi)$ varying from $0$ to some $K_{d,p}>0$ for different $\bxi,\bsi$. The dependence $k_{d,p}=k_{d,p}(\bxi,\bsi)$ should be taken into account in the induction hypotheses which makes things complicated.
\end{remark}
\begin{remark}\lbl{r:constants}
		\textit{(i)} Working under the \textit{subcritical} scaling, in the present paper we do not need to control the dependence of the function $C^\#_{m,q}(s)$ from the r.h.s. of \eqref{bound_a} on $m$ and $q$. This is not the case when working under the \textit{critical} scaling,  when $\eps$ is a fixed small constant, independent from $L$. 
		  In Appendix~\ref{a:const_growth} we show that our method allows to specify bound \eqref{bound_a}  as
		\be\lbl{contr_const_intro}
		\EE\sup_{\tau\le l \le \tau+1}\big|a_s^{(m)}(l)\big|^{2q} \leq 	C_r\lan s\ran^{-2qr}\big(q(2m+1)\big)!\,, \qquad \forall r\ge 0,
	\ee
 uniformly in $\tau$ and $L$. 	
	Proof of this estimate follows lines of the above given sketch but
is slightly trickier.  Namely, we have to run the induction not for the single cumulant $\ka(J)$ but for a combination from the l.h.s. of \eqref{sup_est_num_a}.  
For the sake of brevity in Appendix~\ref{a:const_growth} we establish \eqref{contr_const_intro} without the $\sup$ over  time $l$. The estimate \eqref{contr_const_intro} can be deduced from the established one and similar bounds by a standard argument,  used e.g. in Corollary~\ref{l:bound_high_mom_num}.

\textit{(ii)}	
The main obstacle in extending the result of Theorem~\ref{t:main} to the {\it critical} scaling is the factorial dependence on $m$ of  estimate \eqref{contr_const_intro}. We believe that the latter can be improved by a refinement of the  sketched above proof  since the suggested inductive procedure is rough. Indeed, we do not take into account that $\ka(J) = 0$ unless $\sum_{j=1}^p|\xi_j|^2 = \sum_{j=1}^p|\s_j|^2$.
	\footnote{We are grateful to Alberto Maiocchi for this observation.}
\end{remark}

\subsection{Approximation of the exact solution by the quasisolutions}
Next we are aiming to show that under scaling \eqref{subcr} the quasisolution $A^{M}(\tau)$  is close to the exact solution $a(\tau)$ of eq. \eqref{eff} once $M$ is sufficiently large.
To this end we employ a method in spirit of that used in
\cite{CG1, CG2, DH, DH1} and other works.
First we note that $A^M$ satisfies eq. \eqref{eff} with a disparity, which we will show to be small:
\begin{equation}\label{eff_A}
		\dot A_s^M+\gamma_s A_s^M = i\eps Y_s(A^M)
		+b(s) \dot\beta_s - i\eps^{M+1}R_s^M, \qquad A_s^M(0) = 0
		\,,
\end{equation}
where 
\be\lbl{RRR_s^M}
	R_s^M= \sum_{\substack{m_1,m_2,m_3\le M:\\ m_1+m_2+m_3\ge M}} \eps^{m_1+m_2+m_3-M}
Y_s(a^{(m_1)}, a^{(m_2)},  a^{(m_3)}).
\ee
Then, subtracting equations \eqref{eff} and \eqref{eff_A}, we see that the error term $w^M$ from \eqref{a-A-w} satisfies
\begin{align}\non
	\dot w_s^M+\gamma_s w_s^M = i\eps\Big( Y^{sym}_s(w^M,A^M, A^M) &+ Y^{sym}_s(w^M,w^M, A^M) + Y_s(w^M) \\ \lbl{w-eq}
	 &+ \eps^M R_s^M\Big), \qquad w_s^M(0)=0,
\end{align}
where we recall that $Y^{sym}$ is defined in \eqref{Y^sym}.
 
Let  $\cL=\cL(M)$ be a (random) linear operator, acting on a vector $y = \big\{y_s(\tau): \; s\in\Z^d_L, \, \tau\ge 0\big\}$ with $y_s(\tau)\in\C$  as
\be\lbl{lin_op_def}
y\mapsto \cL y, \quad
(\cL y)_s(\tau) = i\cY_s^{sym}(y, A^M, A^M)(\tau).
\ee
Then, applying the Duhamel formula to eq. \eqref{w-eq}, we get
\be\lbl{main_eq}
(\Id - \eps\cL)w^M = i\eps\Big(\cY^{sym}(w^M,w^M,A^M) + \cY(w^M) + \eps^M\cR^M\Big),
\ee
where  $\cY_s(w^M):= \cY_s(w^M,w^M,w^M)$ and $\cR_s^M(\tau) :=  \int_0^\tau e^{-\ga_s(\tau-l)}\,R^M_s(l)\,dl$, i.e.
$\cR_s^M$ is given by \eqref{RRR_s^M} with the operator $Y_s$ replaced by $\cY_s$.

The following result ensures that, under subcritical scaling \eqref{subcr}, for any $\varrho\ge 0$ and sufficiently large $M=M(\varrho)$  the quasisolution $A^M(\tau)$ is  $\eps^\varrho$-close to the exact solution $a(\tau)$  uniformly in $\tau\in[0,T]$, with overwhelming probability. 
\begin{theorem}\lbl{t:quasi-exact}
	Let $d\ge 3$. Assume scaling \eqref{subcr}, the bound $r_*>d-1$ and take any $\varrho\ge  \al^{-1}(2d+5)$. Then
	for any 
$	 M \ge  \al^{-1}(2d+7) + \varrho,$
any $r\ge 0$	and $L\ge L_{M,r}$ with sufficiently large $L_{M,r}$, there is an event
		 $\Om_L = \Om_L(M, r)$ satisfying
		 \be\lbl{Om_L_prob}
		 \PP (\Om_L) \ge 1- C_{M,r}^\#(L), 
		 \ee
and a random variable $w^M\in\frak X_r$,	such that for any $\om\in\Om_L$
	 
	(1)  $w^{M,\om}$ is  a unique 
	solution to eq. \eqref{w-eq} in the space $\frak X_r$;
	
	(2) the solution $w^{M,\om}$ satisfies
	$
	\big| w^{M,\om} \big|_{\frak X_r} \le \eps^\varrho\, .
	$
\end{theorem}

Theorem~\ref{t:main} almost immediately follows from  Proposition~\ref{t:DKMV-extention} together with Theorem~\ref{t:quasi-exact}.

\subsubsection*{\underline{Scheme of the proof of Theorem~\ref{t:quasi-exact}}}

We start by writing an $n$-th, $n\ge 1$, power of the  linear operator $\cL$ from \eqref{lin_op_def} in the form
\be\non
(\cL^n y)_s(\tau)  = \int_0^\tau \sum_{k\in\Z_L^d}
\Big(\cL_{sk}^{n, +}(\tau,t) y_k(t) 
+ \cL_{sk}^{n, -}(\tau,t) \bar y_k(t) \Big) \, dt,  \qquad n\ge 1,
\ee
where the kernels $\cL_{sk}^{n, \pm}$ can be computed inductively, see Section~\ref{s:est_lin_op}.
Arguing similarly to the proof of Proposition~\ref{l:bound_a}, we show that they  satisfy uniform in $L$ and times $t,\tau$ estimates,  see Propositions~\ref{t:sup_est_oper}, \ref{l:bound_high_mom_oper_sup} and Corollary~\ref{l:oper_mom_est}.  In particular, for $n\ge 2$
\be\lbl{lin_op_est_intro}
\EE |\cL_{sk}^{n, \pm}(\tau,t)|^{2q} \le \lan s \ran^{-2q\epsilon}C_{M,n,q}^\#(s \mp k)\qquad \forall q\in \N,
\ee
 where $\epsilon:=2(r_*-d+1)>0$. 
  We do not control the functions $C_{M,n,q}^\#$ from these estimates  since we do not need it. However, using an argument similar to that from Appendix~\ref{a:const_growth}, it can be shown that they have at most factorial grows in $M,n$ and $q$.
 
In Section~\ref{s:Om} we  define $\Om_L$ as an event in which $|a_s^{(m)}|$ and $|\cL^{n,\pm}_{sk}|$ are of size \footnote{Here $L$  can be replaced by $L^\de$ with any $\de>0$ but then the lower bounds for $\varrho$ and $M$ in Theorem~\ref{t:quasi-exact} will change. }
 $O_s(L)$ 
and decay  sufficiently fast as $|s|\to\infty$ or $ |s  \mp k|\to\infty$ correspondingly, for all $m \le M$ and $2\le n \le N $ with sufficiently large  $M, N\ge 0$. Using the bounds from Proposition~\ref{l:bound_a} and \eqref{lin_op_est_intro} (more precisely, that from Proposition~\ref{l:bound_high_mom_oper_sup}) with increasing $q=q(L)$, we arrive at \eqref{Om_L_prob}. 

Next for $\om\in\Om_L$ we are aiming to invert the operator $(\Id-\eps \cL)$ from the l.h.s. of \eqref{main_eq} in the spaces $\frak X_r$. 
To this end we first observe that the operator norm of $\cL^n:\frak X_r \mapsto \frak X_r$  satisfies $\|\cL^n\|_{\frak X_r} \le C_{r} L^{d+3}$ in $\Om_L$, uniformly in $n\le N$. 
In particular, under scaling \eqref{subcr}, $\|(\eps\cL)^{N}\|_{\frak X_r}\le 1/2$ once $N > \al^{-1}(d+3)$ and $L$ is sufficiently large. Accordingly, $\big\|\big(\Id - (\eps\cL)^{N}\big)^{-1}\big\|_{\frak X_r} \le 2$.
Then we use the presentation
\be\lbl{inv_ident}
(\Id-\eps\cL)^{-1} = \Big(\sum_{k=0}^{N-1}(\eps\cL)^k\Big)\,\Big(\Id - (\eps\cL)^{N}\Big)^{-1},
\ee
previously employed in  \cite{DH} in a related context, leading to the bound 
\[
\big\|(\Id-\eps \cL)^{-1}\big\|_{\frak X_r} \le C_{r} L^{d+3}.
\]
The choice of the event $\Om_L$ immediately implies that the $|\cdot|_{\frak X_r}$-norms of the three terms $\cY^{sym}, \cY$ and $\cR^M$ from the r.h.s. of \eqref{main_eq} are bounded by $C_r L^{d+4}|w^M|_{\frak X_r}^p$ with $p=2,3, 0$ correspondingly. In particular, $|\eps^M\cR^M|_{\frak X_r} \le C_{r}\eps^{\varrho}$ for any $\varrho>0$ once $M=M(\varrho)$ is sufficiently large, so the disparity $\eps^M\cR^M$ is indeed small.

Finally, we consider the (random) mapping on $\frak X_r$ 
\be\lbl{mappinig_intro}
F^M:\; w^M \mapsto (\Id-\eps \cL)^{-1} \Big(\mbox{ r.h.s. of \eqref{main_eq}}\Big).
\ee
The obtained estimates immediately imply that for $\om\in\Om_L$ the mapping $F^{M,\om}$ is a contraction of the $\eps^\varrho$-ball in $\frak X_r$ to itself, once $\varrho$ and $M=M(\varrho)$ are sufficiently large. So $F^{M,\om}$ has a unique fixed point $w^{M,\om}$ in the $\eps^\varrho$-ball, which gives the desired solution to \eqref{w-eq}. A standard argument implies that it is unique not only in the $\eps^\varrho$-ball but in the whole space $\frak X_r$.

\subsubsection*{\underline{Organization of the sequel}}
In Sections~\ref{s:est_a} and \ref{s:est_lin_op} we obtain uniform in $L$ estimates for the terms $a^{(m)}_s$ and the kernels $\cL_{sk}^{n, \pm}$ correspondingly; this is the key part of our paper.
 In Sections~\ref{s:Om}, \ref{s:complete} we prove  Theorem~\ref{t:quasi-exact} and in Section~\ref{s:main_t_proof} ~--- our main result, Theorem~\ref{t:main}.  
In Section~\ref{s:d=2} we discuss the case $d=2$. 
In Appendix~\ref{a:WKI} we introduce the wave kinetic operator $K$ from the WKE \eqref{WKE}. In Appendix~\ref{a:cum} we recall some facts about cumulants. In Appendix~\ref{a:nt} we formulate a number theory result 
which we intensively use when obtaining the estimates from Section~\ref{s:est}. To Appendix~\ref{a:sev_proofs} we postpone some standard proofs or those, almost repeating proofs, given before in the main part of the text. In Appendix~\ref{a:const_growth} we derive the dependence of the functions  $C^\#_{m,q}$ from \eqref{bound_a} on $m$ and $q$.

\ssk
Note that we use the subcritical scaling \eqref{subcr} of $\eps$ only in Sections~\ref{s:complete} and \ref{s:main_t_proof}. Objects, studied in the other sections, are independent from $\eps$.

	\section{Estimates}
\lbl{s:est}

\subsection{Estimates for the terms $a^{(m)}_s$ of  decomposition~\eqref{a-A-w}}
\lbl{s:est_a}

Let us consider a minimal family $\cF$ of random fields $x=\{x_s(\tau):\, s\in\Z^d_L, \, \tau\ge 0\}$, 
that contains the random field $a^{(0)}$
	and is closed with respect to the action of the three-linear operators $Y$ and $\cY$ from \eqref{Ydef} and \eqref{cYdef}.
The set $\cF$ can be constructed  iteratively. Indeed, let $\cF^{(0)}:=\{a^{(0)}\}$ and assume that sets $\cF^{(k)}$ with $k\le n-1$ are already known. Then we define
\be\lbl{cFn}
\cF^{(n)} := \Big\{\Theta(x^1,x^2,x^3):\; \Theta=Y \mbox{ or } \cY,\;\; x^j\in\cF^{(n_j)}, \;\; n_1+n_2+n_3 = n-1\Big\}.
\ee
We set 
$
	\cF:= \bigcup_{n=0}^\infty \cF^{(n)}.
$

	Note that random fields from a set $\cF^{(n)}$ are Wigner chaoses of order $2n+1$.  
Together with the families of random fields  above	we consider  the sets $\bar\cF$ and $\bar\cF^{(n)}$, consisting of the random fields conjugated to those from $\cF$ and $\cF^{(n)}$ correspondingly. 
In particular, $\bar\cF^{(0)} = \{\bar a^{(0)}\}$.

Let us consider a multi-set
	\be\lbl{J_set}
	J = J(\bbeta,\bm\tau)= \Big\{x^1_{\eta_1}(\tau_1), \dots, x^{k}_{\eta_{k}}(\tau_{k})\Big\}, \qquad x^j\in\cF\cup\bar\cF, \qu k\ge 1,
	\ee
	where  $\bbeta=(\eta_1,\dots,\eta_{k})$, $\eta_j\in\Z^d_L$, is a multi-vector of indices and $\bm{\tau} = (\tau_1, \dots, \tau_k)$, $\tau_j\ge 0$, is a vector of times. 
	Denote
	\[
	\deg x^j:=n \qmb{if}\qu x^j\in\cF^{(n)}\cup\bar\cF^{(n)} \qnd \deg J:= \sum_{j=1}^{k} \deg x^j.
	\]
	The main result of this section is Proposition~\ref{t:sup_est_num}, in which, arguing by induction in $\deg J$, we establish a  uniform in $L$ bound for the cumulant $\ka(J)$, viewed as a function of $\bbeta,\bm\tau$. We start by  collecting some basic properties of the latter  	in the lemma below.
		
		Let $I_+(J)$ and $I_-(J)$ be the multi-sets of indices of {\it non-conjugated} and {\it conjugated} random fields from $J$,
	\[
	I_+(J) = \{\eta_j:\; x^j \in \cF\}, \qquad I_-(J)=  \{\eta_j:\; x^j \in \bar\cF\}.	
	\]
	Clearly, $|I_+(J)|+ |I_-(J)|=|J| = k$,
 where we recall Notation~\ref{s:not}(2).
	\begin{lemma}\lbl{l:cum_basics} If $\ka(J) \ne 0$ then
		\begin{enumerate}
			\item
			$|I_+(J)| = |I_-(J)|$;
			\item 
			$\sum I_+(J) = \sum I_-(J)$;
			\item if additionally $\deg J = 0$  then  $k= 2$;
			\item $k$ is even.
		\end{enumerate}
	\end{lemma}
 According to item (4), from now on we always assume 
	\be\lbl{ord_even}
	k=2p, \qquad p\ge 1.
	\ee
 	We will establish the lemma later, simultaneously with the next one, in which  we provide a decomposition for the cumulant $\ka(J)$, crucial for our analysis.
	Let us assume that $\deg x^1\ge 1$ and,  for definiteness, that $x^1$ is non-conjugated , i.e. $x^1\in\cF$. Then
	\be\lbl{x^1=}
	x^1 = \Theta(y^1, y^2, y^3) \qmb{for some}\qu y^j\in \cF \qnd \Theta = Y\mbox{ or }\cY.
	\ee
	Denote 
	\be\lbl{JJ_new}
	\check J := J\sm \big\{x^1_{\eta_1}(\tau_1)\big\} \qnd 
	J_{new} := \Big\{y^1_{s_1}(l), y^2_{s_2}(l), \bar y^3_{s_3}(l)\Big\}\,,
	\ee
	where $l \ge 0$ and $s_j\in\Z^d_L$. Recall that the set of partitions $\cP_2(\check J\sqcup J_{new})$ is defined in Notation \ref{s:not}(3).
	For any $\pi\in \P_2(\check J\sqcup J_{new})$ we denote by $\frak S(\bbeta;\pi)$ the set of all vectors $\bm s = (s_1,s_2,s_3)\in\Z^{3d}_L$, for which
	\be\lbl{frak_S}
	\Del^{s_1 s_2}_{s_3 \eta_1}\ne  0 \qu\qnd\qu \sum I_+(\cA) = \sum I_-(\cA) \qu \forall \cA\in \pi\,.
	\ee
cf. Lemma~\ref{l:cum_basics}(2).
%

	\begin{lemma}\lbl{l:cum_decomp}
		For any multi-set $J$ as in \eqref{J_set} with $x^1$ satisfying \eqref{x^1=},
		\be\lbl{cum_decomp_lem}
		\big|\ka(J)\big| \le \ga_{\eta_1}^{-\mI_\cY(\Theta)}\sum_{\pi\in\P_2(\check J\sqcup J_{new})}
 L^{-d+1}\sum_{\bm s\in \frak S(\bbeta;\pi)} 
\sup_{0\le l\le \tau_1}\prod_{\cA\in\pi} \big|\ka(\cA)\big|\, ,
		\ee
		where $\mI_\cY(\Theta) = 1$ if $\Theta=\cY$ in \eqref{x^1=} and $\mI_\cY(\Theta) = 0$ if $\Theta = Y$.
			\footnote{In this section we do not use the factor $\ga_{\eta_1}^{-\mI_\cY(\Theta)}$ in \eqref{cum_decomp_lem}, but we need it when proving Proposition~\ref{t:sup_est_num_a}, in which we establish analogue of \eqref{k-ind_est} with controlled function $C^\#(\bm\eta)$.}
	\end{lemma}
	Note that the cumulants $\ka(\cA)$ from \eqref{cum_decomp_lem} depends on the multi-vector of indices $\bm s\in\frak S(\bbeta;\pi)$ and time $l\ge 0$ via the elements $y_{s_j}(l)$ of the set $J_{new}$ from \eqref{JJ_new}.
	
	\ssk
	{\it Proofs of Lemmas~\ref{l:cum_basics} and ~\ref{l:cum_decomp}.}
	In \textit{Step 1} below we start to prove bound \eqref{cum_decomp_lem}, it \textit{Step 2} we establish Lemma~\ref{l:cum_basics}, and finally in \textit{Step 3} we conclude the proof of \eqref{cum_decomp_lem}.
	 
	{\it Step 1.} If  $\Theta = \cY$ in \eqref{x^1=}, then, due to the multi-linearity of cumulants and  definition \eqref{Ydef}, \eqref{cYdef} of the operator $\cY$,
	\be
	\non
	\ka(J) =  \int_{0}^{\tau_1}
	e^{-\ga_{\eta_1}(\tau_1-l)} K(l)\, dl \le \ga_{\eta_1}^{-1} \sup\limits_{0\le l\le \tau_1}|K(l)|,
	\ee 
	where
	\be
	\non
	\begin{split}
		K(l)&= \ka\Big(\check J\sqcup\big\{Y_{\eta_1}(y^1,y^2,y^3)(l)\big\}\Big) 
		\\ &=  L^{-d+1} 
		\sum_{s_1,s_2, s_3\in\Z^d_L}\Del^{s_1 s_2}_{s_3 \eta_1} \,\ka\Big(\check J\sqcup \big\{(y^1_{s_1}y^2_{s_2}\bar y^3_{s_3})(l)\big\}\Big).
	\end{split}
	\ee
	If $\Theta = Y$, then
	$
	\ka(J) = K(\tau_1).
	$
	By the Malyshev formula \eqref{Mal},
		\be
	\non
	\ka\Big(\check J\sqcup \big\{(y^1_{s_1}y^2_{s_2}\bar y^3_{s_3})(l)\big\}\Big) = 
	\sum_{\pi\in\P_2(\check J\sqcup J_{new})}
	\prod_{\cA\in\pi} \ka(\cA)\, .
	\ee
	Thus,
	\be\lbl{cum_decomp_non_final}
	|\ka(J)| \le \ga_{\eta_1}^{-\mI_\cY(\Theta)} L^{-d+1}
	\sum_{s_1,s_2, s_3\in\Z^d_L}\Del^{s_1 s_2}_{s_3 \eta_1} \,
	\sum_{\pi\in\P_2(\check J\sqcup J_{new})}
	\sup_{0\le l\le \tau_1} \prod_{\cA\in\pi} |\ka(\cA)|\, .
	\ee
	
{\it Step 2.} 
If $\deg J=0$ then $x^j=a^{(0)}$ or $\bar a^{(0)}$ and are Gaussian, so item (3) follows from \eqref{cum1,2} and \eqref{cum_normal}. Item (4) follows from item (1).

Let us prove items (1) and (2), arguing by induction in $\deg J$. In the Gaussian case $\deg J = 0$ the assertions follow from item (3), \eqref{cum1,2}  and \eqref{a^0-corr}. 
Assume now that $\deg J\ge 1$ and, for definiteness, that \eqref{x^1=} holds, so that \eqref{cum_decomp_non_final} applies.
For any $\pi\in\cP_2(\check J\sqcup J_{new})$, 
\[
\bigsqcup_{\cA\in\pi} I_+(\cA) = \{s_1,s_2\}\sqcup I_+(J)\sm\{\eta_1\}
\qnd 
\bigsqcup_{\cA\in\pi} I_-(\cA) = \{s_3\}\sqcup I_-(J).
\]
Accordingly, 
$\sum_{\cA\in\pi} |I_\pm(\cA)| = |I_\pm(J)| + 1$ and
$$
 \sum_{\cA\in\pi}  \sum I_+(A) = \sum I_+(J) + (s_1+s_2-\eta_1),
\quad
\sum_{\cA\in\pi} \sum I_-(A)  = \sum I_-(J) + s_3\,.
$$
Since $\deg\cA<\deg J$ for every $\cA\in\pi$,  
by the induction hypotheses we have $|I_+(\cA)| = |I_-(\cA)|$ and $\sum I_+(\cA) = \sum I_-(\cA)$.
Since $s_1+s_2-\eta_1 = s_3$ due to the factor $\Del^{s_1s_2}_{s_3\eta_1}$ in \eqref{cum_decomp_non_final}, we get the desired assertions. Proof of Lemma~\ref{l:cum_basics} is completed.

{\it Step 3.} By Lemma~\ref{l:cum_basics}(2) applied to each multi-set $\cA\in\pi$, in \eqref{cum_decomp_non_final} it suffices to take the summation only over those $(s_1,s_2,s_3)$ for which the second relation from \eqref{frak_S} holds. Proof of Lemma~\ref{l:cum_decomp} is completed. 
\qed

\begin{proposition}\lbl{t:sup_est_num}
For any multi-set $J$ as in \eqref{J_set}, \eqref{ord_even},  
	\be\lbl{k-ind_est}
	|\ka(J)| \leq C_{p,\deg J}^\#(\bbeta) L^{-(d-1)(p-1)}\, ,
	\ee
	uniformly in $\bm\tau$.
\end{proposition}
\begin{remark}
	By induction it is straightforward to see that if $p\ge\deg J+2$ then $\ka(J)=0$, cf. Lemma~\ref{l:cum_basics}(3). We do not use and do not prove this fact.
\end{remark}
{\it Proof of Proposition~\ref{t:sup_est_num}.} We argue by induction in $\deg J$. If $\deg J=0$ then the processes $x^j$ are Gaussian, so $\ka(J)\ne 0$ only if $p=1$ and $J=\{a_{s}^{(0)}(\tau_1), \bar a_s^{(0)}(\tau_2)\}$. In this case, due to \eqref{a^0-corr}, we have
\[
|\ka(J)| =\big| \EE a_{s}^{(0)}(\tau_1) \bar a_s^{(0)}(\tau_2)\big| \le \frac{b(s)^2}{\ga_s} = C^\#(s)\, .
\]
Now we assume that  $\deg J\ge 1$ and for definiteness \eqref{x^1=}, so that Lemma~\ref{l:cum_decomp} applies.
By Lemma~\ref{l:cum_basics}(4),  in the r.h.s. of \eqref{cum_decomp_lem} it suffices to take the summation only over those partitions $\pi$ for which $|\cA|$ is even for every $\cA\in\pi$. Then, denoting $p_\cA := |\cA|/2$, by the induction assumption we get
\be\lbl{ka_ind_est}
|\ka(\cA)| \le C_{p_\cA,\deg \cA}^\#(\bm s_\cA, \bbeta_\cA)L^{-(d-1)(p_\cA-1)} \qquad \forall \cA\in\pi,
\ee 
where $\bm s_\cA$ and $\bbeta_{\cA}$ stand for the vectors of indices $s_i$, $\eta_j$ of terms $y^i_{s_i}$, $\bar y^j_{s_j}$, $x^m_{\eta_m}$ from the set $\cA$. Note that 
\be\lbl{paCa}
\sum_{\cA\in\pi} p_\cA = p+1 \qnd \prod_{\cA\in\pi}C_{p_\cA, \deg \cA}^\#(\bm s_\cA, \bbeta_\cA) \le C^\#(\bm s) C^\# (\bbeta_{\ge 2}),
\ee
see Notation~\ref{s:not}(4).
Here and below we use properties of the functions $C^\#$ listed in  Notation~\ref{s:not}(1). Till the end of the proof we do not indicate the dependence of these functions on $p$ and $\deg J$.

Due to \eqref{paCa} and Lemma~\ref{l:cum_decomp},
\be\lbl{K_2*}
|\ka(J)| \le C^\#(\bbeta_{\ge 2})\sum_{\pi\in\P_2(\check J\sqcup J_{new})} 	L^{-(d-1)(p+2-|\pi|)} \,S_\pi(\bbeta) \,, 
\ee
where the function $C^\#$ depends on $p, \,\deg J$ and
\be\lbl{S_pi}
S_\pi(\bbeta) = \sum_{\bm s\in \frak S(\bbeta;\pi)} C^\#(s_1,s_2,s_3)\,.
\ee
Thus, it remains to show that for any $\pi\in\P_2(\check J\sqcup J_{new})$, 
\be\lbl{S_pi-est}
S_\pi(\boldsymbol\eta) \le
 C^\#(\eta_1)   L^{(d-1)(3-|\pi|)} \, .
\ee
Let us substitute 
\be\lbl{subst}
s_1= u+\eta_1 \qnd s_2 = v+\eta_1\, ,
\ee
so that $s_3 = u+v+\eta_1$ (recall that $\de'^{s_1s_2}_{s_3\eta_1}=1$ for $\bm s\in\frak S(\bbeta;\pi)$) and
\be\lbl{Csharp}
C^\#(s_1,s_2,s_3) = C^{\#}(u+\eta_1,v+\eta_1,u+v+\eta_1)\le C_1^\#(\eta_1) C_1^\#(u,v)\, .
\ee
Since 
\be\lbl{om_subst}
\om^{s_1 s_2}_{s_3 \eta_1} =-2u\cdot v = 0 \qmb{for}\qu\bm s\in \frak S(\eta;\pi)
\ee
 (see \eqref{oms}), 
we get
\be\lbl{S_pi_bound}
S_\pi(\bbeta)\le C^\#(\eta_1)\sum_{u,v\in\Z^{d}_L:\, u\cdot v=0}C^\#(u,v) \le  C_1^\#(\eta_1) L^{2(d-1)},  
\ee
due to Corollary~\ref{l:corr_nt} with $a=b=0$.
This is in accordance with 
\eqref{S_pi-est} if $|\pi| = 1$. 

Since $\pi\in\cP_2(\check J \sqcup J_{new})$, each $\cA\in\pi$ satisfies $\cA\cap J_{new} \ne \emptyset$. Then either $|\pi|=1$ or $|\pi|=2,3$, so it remains only to improve the estimate above in the latter case. 
As we will show, to this end it suffices to take into account the second relation in \eqref{frak_S}.

$\bullet$ $a)$ Assume that $|\pi|=2$, so $\pi = \{\cA_1, \cA_2\}$. Consider first the case when 
$ y^1_{s_1}\in\cA_1$ and $y^2_{s_2},  \bar y^3_{s_3} \in\cA_2$.
Due to the second equation in \eqref{frak_S} with $\cA=\cA_1$,  $s_1 = s_1(\bbeta;\pi)$ is  a linear function of $\bm\eta$ once $\bm s\in\frak S(\bm\eta;\pi)$. Then, $u=s_1-\eta_1$ in  \eqref{S_pi_bound} is fixed, so 
the summation is performed only over the vectors $v$ satisfying 
\[
u\cdot v=(s_1(\bbeta;\pi) - \eta_1)\cdot v = 0.
\]
Accordingly, if  $s_1(\bbeta;\pi)\ne \eta_1$, we have $S_\pi(\bbeta) \le C^\#(\eta_1) L^{d-1}$.
If $s_1 = \eta_1$ then  $\de'^{s_1 s_2}_{s_3 \eta_1}\ne 0$ only in the case $s_1=s_2=s_3=\eta_1$, so the summation in \eqref{S_pi_bound} is absent  and $S_\pi\le C^\#(\eta_1)$. Thus,
\be\lbl{K_bar-m^2_num}
S_\pi (\bbeta) 
\le C^\#(\eta_1)L^{d-1},
\ee
in accordance with \eqref{S_pi-est} since $|\pi|=2$.

 $b)$ Due to the symmetry of the indices $s_1$ and $s_2$ it remains to study the case  when 
$ y^1_{s_1}, y^2_{s_2}\in\cA_1$ and $\bar y^3_{s_3} \in\cA_2$.  
By the second relation in \eqref{frak_S} with $\cA=\cA_2$,  $s_3 = s_3(\bbeta;\pi)$ is  a linear function of $\bm\eta$.
Recalling that $u = s_3-\eta_1-v$, we denote $\zeta:= (s_3-\eta_1)/2$ and $w:= v- \zeta\in \Z^d_{L/2}$, so that $u=\zeta-w$ and $u\cdot v =  |\zeta|^2 - |w|^2$. 
Moreover,  
\[
C^\#(u,v) = C^\#(\zeta-w, w+\zeta)\le C^\#_1(\zeta)C^\#_1(w).
\]
Then, by \eqref{S_pi_bound},
\[
S_\pi(\bbeta)\le C^\#(\eta_1)C^\#(\zeta)\sum_{w\in\Z^{d}_{L/2}:\, |w|=|\zeta|} C^\#(w) \le  C_1^\#(\eta_1)L^{d-1},
\]
as in \eqref{K_bar-m^2_num}.

$\bullet$ If $|\pi| = 3$ we have
 $\pi=\{\cA_1,\cA_2,\cA_3\}$ with $|\cA_k\cap J_{new}|=1$. Then, applying again the second relation in \eqref{frak_S} to each set $\cA_k$, we see that the set $\frak S(\bbeta;\pi)$ consists of at most one point $(s_1,s_2,s_3)$  with $s_k=s_k(\bbeta;\pi)$ for $k=1,2,3.$
So, by \eqref{S_pi} and \eqref{Csharp},
$
S_\pi \le C^\#(\eta_1),
$
which is again in agreement with \eqref{S_pi-est}.
\qed

\begin{corollary} \lbl{l:bound_high_mom_num} 
(1)	For any $x\in\cF\cup\bar\cF$ and $p\ge 1$,
	\begin{align}\lbl{bound_high_mom_num}
		\EE  |x_s(\tau)|^{2p} \leq C_{p,\deg x}^\#(s) \qmb{uniformly in}\qu\tau\ge 0,\qu L\ge 1.
	\end{align}

(2) If additionally $x$ is of the form \eqref{x^1=} with $\Theta = \cY$, then
\begin{align}\lbl{bound_high_mom_num_sup}
	\EE \sup_{\tau\le l\le \tau+1} |x_s(l)|^{2p} \leq C_{p,\deg x}^\#(s) \qmb{uniformly in}\qu \tau\ge 0, \qu L\ge 1.
\end{align}
\end{corollary}
{\it Proof.}
(1) Let us consider the set $J$ from \eqref{J_set}, \eqref{ord_even} with $x^j_{\eta_j}(\tau_j):=x_s(\tau)$ for all $1\le j\le p$ and $x^j_{\eta_j}(\tau_j):=\bar x_s(\tau)$ for $p+1\le j\le 2p$.
Then, 
\be\lbl{Ex-cumJ}
\EE |x_s(\tau)|^{2p}= \EE \prod_{z\in J}z  =
 \sum_{\pi\in\P(J)} \prod_{\cA\in\pi} \ka(\cA),
\ee
 according to the Leonov-Shiryaev formula \eqref{mom-cum}. Due to Lemma~\ref{l:cum_basics}(4), $\ka(\cA)\ne 0$ only if $|\cA| = 2p_\cA$ for some $p_\cA\ge 1$. Applying to each cumulant $\ka(\cA)$ Proposition \ref{t:sup_est_num} and using that $\sum_{p_\cA}=p$,  we find 
$$
\EE |x_s(\tau)|^{2p} \le C^\#(s) \sum_{\pi\in\P(J)} L^{-(d-1)(p-|\pi|)} \le C_1^\#(s) ,
$$
since $|\pi|\le p$. Indeed, otherwise at least one of the sets $\cA\in\pi$ must be a singleton which contradicts to $|\cA|=2p_\cA$.

(2) For any $\tau\le l\le \tau+1$,
\be\non
|x_s(l)|^{2p} - |x_s(\tau)|^{2p} =  \int_{\tau}^{l} \frac{d}{dt}|x_s(t)|^{2p}\,dt = 
2p\Re \int_{\tau}^{l} |x_s(t)|^{2p-2} \dot x_s(t) \bar x_s(t)\,dt.
\ee
 Since $x$ is of the form \eqref{x^1=} with $\Theta = \cY$, in view of \eqref{cYdef} we have $\dot x_s = -\ga_sx_s + Y_s(y^1,y^2,y^3)$. Then,
 \be\lbl{sup_ineq}
 \begin{split}
 |x_s(l)|^{2p} - |x_s(\tau)|^{2p} &\le 
 2p \Re \int_{\tau}^{l} |x_s(t)|^{2p-2}   Y_s(y^1,y^2,y^3)(t) \bar x_s(t)\,dt \\
 &\le 2p\int_\tau^{\tau+1} |x_s(t)|^{2p} + |Y_s(y^1,y^2,y^3)(t)|^{2p} \,dt.
\end{split}
 \ee
Since $x\in\cF$ and $Y(y_1,y_2,y_3)\in\cF$, estimate \eqref{bound_high_mom_num} applies to the both  random fields. Then, taking $\EE \sup_{\tau\le l\le \tau+1}$ of the both sides of \eqref{sup_ineq}, we get \eqref{bound_high_mom_num_sup}.

{\bf Proof of Proposition \ref{l:bound_a}.}
Since $a^{(0)}_s$ is the Ornstein-Uhlenbeck process, for $m=0$ the claimed estimate is well-known and immediately follows from the Doob-Kolmogorov inequality and the fact that the function $b(s)$ is fast decaying.
Let now $m>0$.
	Denote by $\cT_m$ the set of all ternary trees with $m$ internal nodes. Due to \eqref{an},
\be\lbl{a---x}
a^{(m)} = i^m\sum_{\cT\in\cT_m} x(\cT), \qquad  x(\cT)\in\cF^{(m)},
\ee
where the terms $x(\cT)$ satisfy assumptions of Corollary~\ref{l:bound_high_mom_num}(2). Then the desired bound follows from \eqref{bound_high_mom_num_sup}.
\qed 

\subsection{Estimates for the linear operator $\cL$}
\lbl{s:est_lin_op}

In this section we study the linear operator $\cL$ defined in \eqref{lin_op_def}, together with its $n$-th degree $\cL^n$, $n\ge 1$. 
Throughout the section we skip the upper index $M$ in the notation $A^M$ for the quasisolution.

\subsubsection*{Kernels} We write $\cL^n$  in the form
\be\lbl{L^n-deff}
(\cL^n y)_s(\tau)  = \int_0^\tau \sum_{k\in\Z_L^d}
\Big(\cL_{sk}^{n, +}(\tau,l	) y_k(l) 
+ \cL_{sk}^{n, -}(\tau,l) \bar y_k(l) \Big) \, dl, 
\ee
where $y = \{y_s(\tau):\; s\in\Z^d_L, \, \tau\ge 0\}$,
\be\lbl{lin_ker_1}
\cL_{sk}^{1, +}(\tau,l) = 2i e^{-\ga_s(\tau-l)}\, L^{-d+1} \sum_{s_1,s_3\in\Z^d_L} \Del^{s_1k}_{s_3s} A_{s_1}(l) \bar A_{s_3}(l),
\ee
$A=A^M$
and
\be\lbl{lin_ker_2}
\cL_{sk}^{1, -}(\tau,l) = i e^{-\ga_s(\tau-l)}\, L^{-d+1} \sum_{s_1,s_2\in\Z^d_L} \Del^{s_1s_2}_{k\,s} A_{s_1}(l) A_{s_2}(l).
\ee
Then the kernels $\cL_{sr}^{n, \zz}$ with $\zz = \{+,-\}$ and $n\ge 2$ are computed iteratively  as
\be\lbl{lin_ker}
\cL_{sr}^{n, \fz}(\tau,t) = \int_t^\tau\,  \sum_{k\in\Z^d_L}\Big(\cL_{sk}^{1,+}(\tau,l) \cL_{kr}^{n-1, \fz}(l,t) 
+ \cL_{sk}^{1,-}(\tau,l) \ov {\cL_{kr}^{n-1, -\fz}}(l,t)\Big)\, dl \, .
\ee
It is convenient to set
\be\lbl{LW^0}
\cL_{sk}^{0, +}(\tau,t)  := \de_{sk} \qnd \cL_{sk}^{0, -} := 0,
\ee
where $\de_{sk}$ is the Kronecker delta. Then definitions \eqref{lin_ker_1} and \eqref{lin_ker_2} take the form
\be\lbl{lin_ker_1'}
\cL_{sk}^{1, +}(\tau,t) = 2i e^{-\ga_s(\tau-t)}\, L^{-d+1} \sum_{s_1,s_2,s_3} \Del^{s_1s_2}_{s_3s} A_{s_1}(t) \,\cL^{0,+}_{s_2k}\,\bar A_{s_3}(t)
\ee
and
\be\lbl{lin_ker_2'}
\cL_{sk}^{1, -}(\tau,t) = i e^{-\ga_s(\tau-t)}\, L^{-d+1} \sum_{s_1,s_2,s_3} \Del^{s_1s_2}_{s_3s} A_{s_1}(t) A_{s_2}(t)\, \ov{\cL^{0,+}_{s_3k}}.
\ee

\subsubsection*{The multi-set $\cJ$} Our goal is to get uniform in $L$ estimates for the kernels $\cL^{n,\pm}_{sr}$. To this end we employ argument very similar to that used in Section~\ref{s:est_a} but more cumbersome because of notational difficulties. 
We consider a multi-set 
\be\lbl{JJ}
\cJ = \big\{X^1_{\xi_1 \s_1}, \dots, X^m_{\xi_m \s_m}, x^1_{\eta_1}, \dots, x^{k}_{\eta_{k}}\big\},
\ee
where  $\xi_i,\s_j,\eta_r\in \Z^d_L$, \; $x^j\in\{A(\tau_j'), \bar A(\tau_j')\}$, $\tau_j'\ge0$, and 
\[
X^j\in\Big\{\cL^{n_j,\zz_j}(\tau_j, t_j),\, \ov{\cL^{n_j,\zz_j}}(\tau_j, t_j)  \Big\},
\]
with $\zz_j\in\{+,-\}$, $0\le t_j\le \tau_j$ and $n_j \ge 0$. We denote $\bm\xi = (\xi_1,\dots,\xi_m)$ and similarly define the vectors $\bm\s, \bm\eta, \bm\zz$ and $\bm\tau,\bm t, \bm \tau'$.
 
 \subsubsection*{Signs of indices} Let us assign signs to the indices $\xi_i,\s_j,\eta_r$ by the rule
 \be\lbl{X=W}
 \mbox{if}\qu X_{\xi_j\s_j}^j = \cL_{\xi_j\s_j}^{n_j,\zz_j} \qu\Rightarrow\qu \sign\xi_j = +, \; \sign\s_j = -\zz_j
 \ee
 and for $x_{\eta_j}^j = A_{\eta_j}$ we set $\sign\eta_j = +$. Conjugation invert signs of the indices: for $X_{\xi_j\s_j}^j = \ov{\cL_{\xi_j\s_j}^{n_j,\zz_j}}$ we define $\sign\xi_j = -, \; \sign\s_j = \zz_j $ and for $x_{\eta_j}^j = \bar A_{\eta_j}$ we set $\sign\eta_j = -$.
 We introduce the multi-sets $I_+(\cJ)$ and $I_-(\cJ)$ of indices with positive and negative signs as 
 $$
 I_\pm(\cJ) = \big\{s\in\{\xi_i,\s_j,\eta_r\}:\; \sign s = \pm\big\}.
 $$
 Clearly, $|I_+(\cJ)| + |I_-(\cJ)| =|\cJ|= 2m+k$.

 \subsubsection*{Properties of cumulants $\ka(\cJ)$} 
 The result below is analogous to items (1,2,4) of Lemma~\ref{l:cum_basics}. 
 \begin{lemma}\lbl{l:cum_basics_L}
 	If $\ka(\cJ) \ne 0$ then
 	\ssk
 	
 	(1)\; $|I_+(\cJ)| = |I_-(\cJ)|$;
 	\quad (2)\; $\sum I_+(\cJ) = \sum I_-(\cJ);$
 	\quad (3)\;  $k$ is even.
 \end{lemma}
 In view of Lemma~\ref{l:cum_basics_L}(3), from now on we always assume that $k=2p$ with $p\ge 1$, as in \eqref{ord_even}.
 
We prove Lemma~\ref{l:cum_basics_L} in Appendix~\ref{a:cum_prop_op}, repeating the argument used to establish Lemma~\ref{l:cum_basics}. 
The proof, as well as the main estimate of this section stated below, is based on the following decomposition, analogous to that from Lemma~\ref{l:cum_decomp}.
%
Let us assume that
\be\lbl{for_definit}
X^1 = \cL^{n_1,\zz_1} \qmb{with} \qu n_1 \ge 1, \qu \zz_1 \in\{+,-\}\, ,
\ee
so that for $n_1\ge 2$ the kernel $X^1$ decomposes via \eqref{lin_ker_1}-\eqref{lin_ker}, while for $n_1=1$ it has the form \eqref{lin_ker_1} or \eqref{lin_ker_2}.
Consider multi-sets
\be\lbl{J_new_pm}
\check \cJ := \cJ\sm X^1_{\xi_1\s_1}, \qquad \cJ_{new}^+ := \big\{A_{s_1}(l), \,\cL_{s_2\s_1}^{n_1-1,\zz_1}(l,t_1), \,\bar A_{s_3}(l) \big\}
\ee
and
\be\lbl{J_new_-}
\cJ_{new}^- := \big\{A_{s_1}(l), \, A_{s_2}(l),\, \ov{\cL_{s_3\s_1}^{n_1-1,-\zz_1}}(l,t_1)\big\},
\ee
cf. \eqref{lin_ker_1'}, \eqref{lin_ker_2'}.
For any $\pi\in \P_2(\check \cJ\sqcup \cJ^\pm_{new})$ we denote by $\frak S(\bxi,\bsi,\bbeta;\pi)$ the set of all vectors $\bm s = (s_1,s_2,s_3)\in\Z^{3d}_L$, for which $\Del^{s_1 s_2}_{s_3 \xi_1}\ne  0$ and $\ka(\cA)\ne 0$ for all $\cA\in\pi$. In particular, by  Lemma~\ref{l:cum_basics_L}(2),
\be\lbl{frak_S_L}
\sum I_+(\cA) = \sum I_-(\cA) \qquad \forall \cA\in \pi\qnd \forall \bm s\in\frak S(\bxi,\bsi,\bbeta;\pi)\, .
\ee
	\begin{lemma}\lbl{l:cum_decomp_L}
	For any multi-set $\cJ$ as in \eqref{JJ}, \eqref{for_definit},
	\be\lbl{cum_decomp_lem_L}
	\big|\ka(\cJ)\big| \le 2\ga_{\xi_1}^{-\mI_{n_1\ge 2}}\,\sum_{\zz=\pm}\sum_{\pi\in\P_2(\check \cJ\sqcup \cJ^\zz_{new})}
	L^{-d+1}\sum_{\bm s\in \frak S(\bxi,\bsi,\bbeta;\pi)} 
	\sup_{t_1\le l\le \tau_1}\prod_{\cA\in\pi} \big|\ka(\cA)\big|\, ,
	\ee
	where $\mI_{n_1\ge 2} := 1$ if $n_1\ge 2$  and $\mI_{n_1\ge 2}  := 0$ otherwise.
\end{lemma}
{\it Proof.}
If $n_1\ge 2$, then, decomposing  $X^1$ via \eqref{lin_ker_1}-\eqref{lin_ker},  we get
\be\non
 \ka(\cJ) =  \int_{t_1}^{\tau_1}
ie^{-\ga_{\xi_1}(\tau_1-l)} \cK(l)\, dl, 
\ee 
where 
\be\lbl{KKK}
\begin{split}
\cK(l):= 
L^{-d+1} 
\sum_{s_1,s_2,s_3}
\Del^{s_1 s_2}_{s_3 \xi_1}\,\Big[  
2&\ka\Big(\check \cJ\sqcup \big\{A_{s_1}\cL_{s_2\s_1}^{n_1-1,\zz_1} \bar A_{s_3} \big\}\Big)
\\
&+ 
\ka\Big(\check \cJ\sqcup \big\{A_{s_1} A_{s_2} \ov{\cL_{s_3\s_1}^{n_1-1,-\zz_1}} \big\}\Big)
\Big],
\end{split}
\ee
 $A_{s_j} = A_{s_j}(l)$ and $\cL_{s_j\s_1}^{n_1-1,\pm} = \cL_{s_j\s_1}^{n_1-1,\pm}(l,t_1)$.
Then, by the Malyshev formula~\eqref{Mal},
\be\lbl{lin_op_K}
\cK(l)= 
L^{-d+1} 
\sum_{s_1,s_2,s_3}\Del^{s_1 s_2}_{s_3 \xi_1}\Big(  
2\sum_{\pi\in\P_2(\check \cJ\sqcup \cJ_{new}^+)}
+ \sum_{\pi\in\P_2(\check \cJ\sqcup \cJ_{new}^-)}\Big)
\prod_{\cA\in\pi} \ka(\cA)\, .
\ee
If $n_1=1$,  due to \eqref{lin_ker_1'}, \eqref{lin_ker_2'},
\be\lbl{lin_op_integr_W}
\ka(\cJ) =   ie^{-\ga_{\xi_1}(\tau_1-t_1)}\cK(t_1).
\ee 
Accordingly, for any $n_1\ge 1$,
\be\lbl{ka_J-est_L}
|\ka(\cJ)|\le 
2 \ga^{-\mI_{n_1\ge 2}}_{\xi_1}L^{-d+1} 
\sum_{s_1,s_2,s_3}\Del^{s_1 s_2}_{s_3 \xi_1} \sum_{\zz=\pm}\sum_{\pi\in\P_2(\check \cJ\sqcup \cJ_{new}^\zz)}
\sup_{t_1\le l\le \tau_1}
\prod_{\cA\in\pi} |\ka(\cA)|\, .
\ee
\qed

\subsubsection*{Estimates} 
Let us set
\be\lbl{GaJ}
\epsilon := 2(r_* - d + 1) >0 \qnd \Ga_\cJ:= \prod\limits_{j=1}^m \lan\xi_j\ran^{\epsilon\mI_{n_1\ge 2}}\, .
\ee
Denote $\bm\zz\odot\bm\sigma:= (\zz_1\s_1,\dots,\zz_m\s_m)$ and $\cN:= \sum_{j=1}^m n_j.$
Recall that $M$ is the order of the quasisolution $A=A^M$.
\begin{proposition} \lbl{t:sup_est_oper}
	For any 	multi-set $\cJ$ as in \eqref{JJ},  \eqref{ord_even} 
	\be\lbl{sup_est_num_L}
	|\ka(\cJ)| \le  \Ga_\cJ^{-1}\, C^\#(\bxi - \bm\zz\odot\bsi) C^\#(\bbeta)  L^{-(d-1)(p + m -1)}, 
	\ee
	where the function $C^\#$ depends on $p, m, M$ and $\cN$.
\end{proposition}
{\it Proof.} 
We argue by induction in $\cN$. 
 If $\cN=0$ then, due to \eqref{LW^0},  the variables $X^j_{\xi_j\s_j}$ are not random and have the form $X_{\xi_j\s_j}^j = \de_{\xi_j\s_j}\de_{\zz_j+}$. Then $\ka(\cJ)$ must vanish unless $m=0$ or $m=1$, $p=0$ (see properties of cumulants listed in Appendix~\ref{a:cum}).
 If $m=0$ the desired estimate follows from Proposition~\ref{t:sup_est_num} since the quasisolution $x^j=A^M$ is a finite linear combination of elements from the set $\cF$ (defined in the beginning of Section~\ref{s:est_a}), according to \eqref{a---x}. In the case   $m=1$, $p=0$ we have $\ka(\cJ)=\de_{\xi_j\s_j}\de_{\zz_j+}$, so  the assertion is again true. 

Let us now assume that the statement is proven for any $\cJ$ with $\cN\le N$ and establish it for $\cJ$ with  $\cN=N+1$. We assume for definiteness \eqref{for_definit} and argue as when proving Proposition~\ref{t:sup_est_num}.
Due to Lemma~\ref{l:cum_decomp_L}, the induction assumption and the multiplicativity of the function $\Ga_\cJ$,
\be\lbl{k(J)<1}
\begin{split}
|\ka(\cJ)|  \le 
	\ga_{\xi_1}^{-\mI_{n_1\ge 2}}C^\#(\bxi_{\ge 2} - \bm\zz_{\ge 2}\odot\bsi_{\ge 2}) &C^\#(\bbeta) 
\sum_{\zz = \pm}
\Ga_{\check \cJ\sqcup \cJ^\zz_{new}}^{-1}
\sum_{\pi\in\P_2(\check \cJ\sqcup \cJ_{new}^\zz)} 
\\&
L^{-(d-1)(p+2+m-|\pi|)} S_\pi^\zz,
\end{split}
\ee
 where
\be\lbl{lin_S_pi}
S^+_{\pi}(\bxi,\bsi,\bbeta) = \sum_{(s_1,s_2,s_3)\in \frak S(\bxi,\bsi,\bbeta;\pi)}  C^\#(s_2-\zz_1 \s_1)C^\#(s_1, s_3)
\ee
and 
\be\lbl{lin_S_pi'}
S^-_{\pi}(\bxi,\bsi,\bbeta) = \sum_{(s_1,s_2,s_3)\in \frak S(\bxi,\bsi,\bbeta;\pi)}
 C^\#( s_3 +\zz_1 \s_1) C^\#(s_1, s_2) \,.
\ee
Since $\Ga_{\check \cJ\sqcup \cJ^\zz_{new}} \ge 	\lan\xi_1\ran^{-\epsilon\mI_{n_1\ge 2}} \Ga_\cJ$ and
$\ga_{\xi_1}\ge C^{-1}\lan \xi_1\ran^{2r_*}$, 
\be\lbl{gaGa}
\ga_{\xi_1}^{-\mI_{n_1\ge 2}} \Ga_{\check\cJ \sqcup \cJ_{new}^\zz}^{-1} 
\le
C \lan\xi_1\ran^{(-2r_* + \epsilon)\mI_{n_1\ge 2}}\Ga_\cJ^{-1} = C \lan\xi_1\ran^{2(-d+1)\mI_{n_1\ge 2}}\Ga_\cJ^{-1} .
\ee
Then
it remains to show that 
\be\lbl{des_est_lin}
S_\pi^\pm\le \lan\xi_1\ran^{2(d-1)\mI_{n_1\ge 2}}C^\#(\xi_1-\zz_1\s_1) L^{(d-1)(3-|\pi|)}.
\ee
 Let us substitute 
 \[
 s_1= u+\xi_1 \qnd s_2 = v+\xi_1\, ,
 \]
 so that $s_3 = u+v+\xi_1$ once $\de'^{s_1s_2}_{s_3\xi_1}=1$. Then,
 \begin{align}\non
 C^\#(s_2-\zz_1 \s_1)C^\#(s_1, s_3)  &= C^\#(v+\xi_1 - \zz_1\s_1) C^\#(u+\xi_1, u+v+\xi_1)\\
 \lbl{C-gf1}
 &\le C_1^{\#}(\xi_1-\zz_1 \s_1) C_1^\#(u + \xi_1,v)\,,
 \end{align}
 since $(\xi_1-\zz_1\s_1, u+\xi_1,v)\leftrightarrow(v+\xi_1-\zz_1\s_1, u+\xi_1,u+v+\xi_1)$ is a linear isomorphism (see {\it Properties of functions $C^\#$} in Notation~\ref{s:not}(1)).
 Similarly,
 \begin{align}\non
  C^\#( s_3 +\zz_1 \s_1) C^\#(s_1, s_2)   &= C^\#(u+v+\xi_1 +  \zz_1\s_1)C^\#(u+\xi_1, v+\xi_1) \\
  \lbl{C-gf2}
  &\le C_1^{\#}(\xi_1-\zz_1 \s_1) C_1^\#(u + \xi_1,v + \xi_1)\, .
 \end{align}
 Then, recalling that
$\om^{s_1 s_2}_{s_3 \xi_1} =-2u\cdot v$  (see \eqref{oms}), 
we get
\be\lbl{S_pi_bound_L}
S^\zz_\pi\le  C^{\#}(\xi_1-\zz_1 \s_1)\sum_{u,v\in\Z^{d}_L:\, u\cdot v=0}C_1^\#(u + \xi_1,v+\xi_1\de_{\zz\, -})\, .
\ee
 Due to the number theory bound from Corollary~\ref{l:corr_nt}, 
 \be\lbl{for_ln}
S^\zz_\pi\le  L^{2(d-1)} C^{\#}(\xi_1-\zz_1 \s_1)\lan\xi_1\ran^{2(d-1)}\,,
 \ee
 which is in accordance with 
 \eqref{des_est_lin} if $|\pi| = 1$ and $n_1\ge 2$. 
 If $n_1=1$ then the variables $\cL^{n_1-1,\pm}_{s_j\s_1}$ from \eqref{J_new_pm} and \eqref{J_new_-}
 are not random, so $\frak S(\bxi,\bsi,\bbeta;\pi)=\emptyset$  once $|\pi|=1$ and accordingly $S^\pm_\pi=0$. Indeed, in this case  we have $\pi=\{\check \cJ\sqcup \cJ_{new}^\zz\}$ but $\ka(\check \cJ\sqcup \cJ_{new}^\zz)=0$ since the set $\check \cJ\sqcup \cJ_{new}^\zz$ contains the constant $\cL^{0,\pm}_{s_j\s_1}$.
Thus, it remains to improve the estimates in the cases $|\pi|=2,3$. 
 
 $\bullet$ Let $\pi = \{\cA_1, \cA_2\}\in \cP_2(\check \cJ\sqcup \cJ_{new}^\zz)$.  
Assume that  $ A_{s_1}\in\cA_1$ and  $\cJ_{new}^\zz \sm \{A_{s_1}\} \subset \cA_2$. The other cases with $|\pi|=2$ bare similar and considered in Appendix~\ref{a:S_pi}.
 Due to eq. \eqref{frak_S_L} with $\cA=\cA_1$,  $s_1 = s_1(\bxi,\bsi,\bbeta;\pi)$ is a linear function of $\bxi,\bsi,\bbeta$ once $\bm s\in\frak S(\bxi,\bsi,\bbeta;\pi)$.  So, in \eqref{S_pi_bound_L} $u=s_1-\xi_1$ is fixed and
 the summation  is performed only over vectors $v$ from the hyperplane 
 \[
 u\cdot v=(s_1(\bxi,\bsi,\bbeta;\pi) - \xi_1)\cdot v = 0.
 \]
 Accordingly, if  $s_1(\bxi,\bsi,\bbeta;\pi)\ne \xi_1$, we have $S^\zz_\pi\le  C^{\#}(s_1-\zz_1 \s_1) L^{d-1}$.
 If $s_1 = \xi_1$ then  $\de'^{s_1 s_2}_{s_3 \xi_1}\ne 0$ only in the case $s_1=s_2=s_3=\xi_1$, so the summation in \eqref{lin_S_pi}, \eqref{lin_S_pi'} is absent  and $S_\pi^\zz\le C^{\#}(\xi_1-\zz_1 \s_1)$, due to \eqref{C-gf1}, \eqref{C-gf2}. This is in accordance with \eqref{des_est_lin}
with $|\pi|=2$.

 $\bullet$ If $|\pi| = 3$ we have
 $\pi=\{\cA_1,\cA_2,\cA_3\}$ with $|\cA_k\cap \cJ^\zz_{new}|=1$. Then,  by \eqref{frak_S_L}, the set $\frak S(\bxi,\bsi,\bbeta;\pi)$ consists of a unique point $(s_1,s_2,s_3)$  with $s_k=s_k(\bxi,\bsi,\bbeta;\pi)$ for $k=1,2,3.$
 So,
 $
 S_\pi^\pm \le C^{\#}(\xi_1-\zz_1 \s_1),
 $
 which is again in agreement with \eqref{des_est_lin}.
\qed

\begin{corollary}\lbl{l:oper_mom_est} For any $M\ge 1$, $n\ge 2$ and $q\ge 1$, 
	\be\lbl{LLL_est}
	\EE |\cL_{sk}^{n,\pm}(\tau,t)|^{2q} \leq 
	\lan s \ran^{-2q\epsilon}C^\#(s \mp k),
	\ee
	uniformly in $0\le t\le\tau$ and $L\ge 1$,  where $C^\# = C^\#_{M,n,q}$.
\end{corollary}
{\it Proof.} We argue as when proving estimate \eqref{bound_high_mom_num}.
Let us take $\cJ$ as in \eqref{JJ} with $k=0$, $m=2q$, $X_{\xi_1\s_1}^1 = \dots = X^q_{\xi_q\s_q} = \cL^{n,\pm}_{sk}$  and
$X^{q+1}_{\xi_{q+1}\s_{q+1}} = \dots = X^{2q}_{\xi_{2q}\s_{2q}} = \ov{\cL^{n,\pm}_{sk}}$.
Due to the Leonov-Shiryaev  formula \eqref{mom-cum}, we find
\begin{align}\non
	\EE |\cL_{sk}^{n,\pm}(\tau,t)|^{2q} 
	&\le 
	\sum_{\pi\in\P(\cJ)} \prod_{\cA\in\pi} |\ka(\cA)|
	\\\non
	&\le	\lan s \ran^{-2q\epsilon} C^\#(s\mp k) \sum_{\pi\in\P(\cJ)} L^{-(d-1)(2q-|\pi|)}, 
\end{align}
according to Proposition~\ref{t:sup_est_oper}. Since $|\pi|\le |\cJ| = 2q$, we get the claimed bound. 
\qed

\begin{proposition} \lbl{l:bound_high_mom_oper_sup} 
	For any $M\ge 1$, $n\ge 2$ and $q\ge 1$, 
	\begin{align}\lbl{bound_high_mom_num_oper}
		\EE \sup\limits_{\tau\le l\le \tau+1}|\cL_{sk}^{n,\pm}(l,t)|^{2q} \leq 
	\lan s \ran^{2(d-1)-(2q-1)\epsilon}\,C^\#( s \mp k )
	\end{align}
	uniformly  $0\le t\le\tau$ and $L\ge 1$, where $C^\# = C^\#_{M,n,q}$.
\end{proposition}
{\it Proof.} 
Due to \eqref{lin_ker}, for $n\ge 2$
$$
\frac{d}{d\tau} 	\cL_{sk}^{n,\pm}(\tau,t) = -\ga_s	\cL_{sk}^{n,\pm}(\tau,t)  	+ \cW_{sk}^{n,\pm}(\tau, t),
$$
where
\be\lbl{W_def}
\cW_{sk}^{n, \fz}(\tau,t) =   \sum_{r\in\Z^d_L}\Big(\cL_{sr}^{1,+}(\tau,\tau) \cL_{rk}^{n-1, \fz}(\tau,t) 
+ \cL_{sr}^{1,-}(\tau,\tau) \ov {\cL_{rk}^{n-1, -\fz}}(\tau,t)\Big)\, .
\ee
Then for, $t\le \tau \le l \le \tau+1$,
\begin{align}\non
	|\cL_{sk}^{n,\pm}(l,t)|^{2q} &- |\cL_{sk}^{n,\pm}(\tau,t)|^{2q} = \int_\tau^l \frac{d}{d {l'}}|\cL_{sk}^{n,\pm}(l',t)|^{2q}\, dl'
	\\\non
	&\le 2q\int_\tau^{\tau+1} |\cL_{sk}^{n,\pm}(l',t)|^{2q-1} |\cW_{sk}^{n,\pm}(l', t)|\,dl'\,.
\end{align}
Accordingly, due to the Cauchy-Bunyakovsky inequality,
\begin{align*}
	\EE \sup\limits_{\tau\le l\le \tau+1}|\cL_{sk}^{n,\pm}(&l,t)|^{2q}
	\le \EE |\cL_{sk}^{n,\pm}(\tau,t)|^{2q} \\
	&+ 2q\int_\tau^{\tau+1} \Big(\EE|\cL_{sk}^{n,\pm}(l',t)|^{4q-2}\Big)^{1/2} \Big(\EE|\cW_{sk}^{n,\pm}(l', t)|^2\Big)^{1/2} \,dl'\,.
\end{align*}
Now the assertion of the proposition follows from Corollary~\ref{l:oper_mom_est} and the estimate
\be\lbl{W_est}
\EE |\cW_{sk}^{n,\pm}(\tau,t)|^{2} \leq 
\lan s \ran^{4(d-1)}C_{M,n}^\#(s \mp k), \qmb{uniformly in $0\le t\le\tau$}
\ee
and $L\ge 1$,
which we establish in Appendix~\ref{a:W_est}. 
To get \eqref{W_est} we write
\be\lbl{ka-2mom}
\EE\big|\cW^{n,\pm}_{sk}\big|^2 =\ka\Big(\big\{\cW^{n,\pm}_{sk}, \ov{\cW^{n,\pm}_{sk}}\big\}\Big) - \Big|\ka\Big(\big\{\cW^{n,\pm}_{sk}\big\}\Big)\Big|^2
\ee
and estimate the cumulants  $\ka\big(\{\cW\}\big)=\EE \cW$ and $\ka\big(\{\cW,\bar \cW\}\big)$ by once or twice applying formula \eqref{W_def} and then using bound \eqref{sup_est_num_L}, almost literally repeating the argument used in the proof of Proposition~\ref{t:sup_est_oper}. 
Note that in bound \eqref{LLL_est} there is no growing with $|s|$ factor  in difference with  \eqref{W_est}, since at each step of our inductive procedure the integration in time in \eqref{lin_ker} gives an additional factor $\ga_s^{-1} < C\lan s \ran^{-2(d-1)}$.
\qed

\section{Proofs of Theorems~\ref{t:quasi-exact} and \ref{t:main}}

\lbl{s:theorems}

In Section~\ref{s:Om} we construct the event $\Om_L$ from Theorem~\ref{t:quasi-exact}. 
In  Section~\ref{s:complete} we complete the proof of Theorem~\ref{t:quasi-exact} and in Section~\ref{s:main_t_proof}~--- that of Theorem~\ref{t:main}.

\subsection{Construction of the "good" event $\Om_L^0$ }
\lbl{s:Om}

For  
$M\ge 1$, $N \ge 2$ and $r\ge  0$ 
let  $\Omega_L^0 = \Om_L^0(M,N,r)$ be the event in which 
\be\lbl{Om-a}
\sup\limits_{\tau\in [0,T]} |a_s^{(m)}(\tau)| \le L\lan s \ran^{-r}, \qquad \forall s\in\Z^d_L \qmb{and}\qu \forall0\le m\le M,
\ee
and
\be\lbl{Om-cL}
\int_0^T \sup_{\tau\in[t,T]} \big| \cL_{sk}^{n,\pm}(\tau,t)\big|\, dt \le L\lan s\mp k\ran^{-r},\qquad  \forall s,k\in\Z^d_L, \qu \forall 2\le n\le N.
\ee
The kernels $\cL_{sk}^{n,\pm}$ considered in this section are built from the quasisolution $A=A^M$ of the order $M$ (recall that the latter enters the definition of the kernels via \eqref{lin_ker_1}, \eqref{lin_ker_2}).
In the next section we will choose the event $\Om_L(M,r)$ from Theorem~\ref{t:quasi-exact} in the form \eqref{N=}.
\begin{proposition}\lbl{l:Om_L-est}
	The event $\Om_L^0 = \Om_L^0(M,N,r)$ satisfies 
	$$\PP(\Om_L^0)\ge 1-  C_{M,N,r}^\#(L)  \,.$$
\end{proposition}
Assertion of the proposition follows from the two lemmas below and is proven in the end of section. The lemmas are obtained by a straightforward application of the Markov inequality combined with the estimates of Propositions~\ref{l:bound_a} and \ref{l:bound_high_mom_oper_sup} with appropriately chosen power $q$.

\begin{lemma}\lbl{l:Om-a}
	For any $m\ge 0$, $r\ge 0$ and $Q\ge 1$, 
	\be\lbl{Om-a-est}
	\PP\big(\sup\limits_{\tau\in [0,T]} |a_s^{(m)}(\tau)| \ge  Q \lan s \ran^{-r}\big) \le C_{m,r}^\#(Q)\lan s \ran^{-(d+1)} \,,
	\ee
	uniformly in $L\ge 1$.
\end{lemma}
{\it Proof.} Let us denote the l.h.s. of \eqref{Om-a-est} by $P$. 
Using the Markov inequality together with Propositions~\ref{l:bound_a}, we find
\be\lbl{Ple}
P \le C_{q,m}^\#(s)\lan s\ran^{2qr} Q^{-2q} \le C_{q,m,r}\lan s\ran^{-(d+1)} Q^{-2q}\, , \qquad \forall q\in\N.
\ee
Let us choose $q=q_{m,r}(Q)\to\infty$ as $Q\to\infty$ so slowly that $C_{q(Q),m,r}\le Q$. Since $Q^{-2q(Q)}=C_{m,r}^\#(Q)$, we get 
\[
P \le Q\lan s\ran^{-(d+1)}C_{m,r}^\#(Q) = C'^\#_{m,r}(Q)\lan s\ran^{-(d+1)}.
\]
\qed

\begin{lemma}\lbl{l:Om-cL-est}
	For any $n\ge 2$,  $r\ge 0$ and $Q\ge 1$,
	\be\lbl{Om-cL-est}
		\PP\Big(\int_0^T \sup_{\tau\in[t,T]} \big| \cL_{sk}^{n,\pm}(\tau,t)\big|\, dt \ge Q \lan s\mp k\ran^{-r} \Big) \le C_{M,n,r}^\#(Q)\Big(\lan s\ran \lan s\mp k \ran\Big)^{-(d+1)}\, ,
	\ee
	uniformly in $L\ge 1.$
\end{lemma}
{\it Proof.} 
Due to the Jensen inequality, for any $q\in\N$ 
\be\lbl{exp_int}
\EE\Big(\int_0^T \sup_{\tau\in[t,T]} \big| \cL_{sk}^{n,\pm}(\tau,t)\big|\, dt\Big)^{2q} \le 
T^{2q-1} \int_0^T \EE\sup_{\tau\in[t,T]} \big| \cL_{sk}^{n,\pm}(\tau,t)|^{2q} \, dt.
 \ee
Then, by the Markov inequality together with Proposition~\ref{l:bound_high_mom_oper_sup}, 
\begin{align}\non
\mbox{l.h.s. of \eqref{Om-cL-est}}
&\le \lan s \ran^{2(d-1) - (2q-1)\epsilon} C_{q,M,n,r}^\#(s\mp k)\lan s\mp k \ran^{2qr} Q^{-2q} \\ \lbl{lhs45}
&\le C_{q,M,n,r}\,\Big(\lan s \ran\lan s\mp k \ran\Big)^{-(d+1)}Q^{-2q} \,,
\end{align}
once $q$ is so large that $2(d-1) - (2q-1)\epsilon\le 0$.
Then, choosing $q=q_{M,n,r}(Q)\to\infty$ so slowly that $C_{q(Q),M,n,r}\le Q$, we get the assertion.
\qed

\ssk

\underline{\it Proof of Proposition~\ref{l:Om_L-est}.} 
To estimate the probability $\PP(\Om\sm\Om_L^0)$ we sum up in $s,k\in\Z^d_L$ bounds for probabilities of the complementary events to \eqref{Om-a} and \eqref{Om-cL}, following from Lemmas~\ref{l:Om-a} and \ref{l:Om-cL-est} with $Q=L$. We get
\be\non
\begin{split}
\PP(\Om\sm\Om_L^0) &\le C_{M,n,r}^\#(L)  \Big(\sum_{s\in\Z^d_L} \lan s\ran^{-(d+1)} + \sum_{s,k\in\Z^d_L} \lan s \ran^{-(d+1)} \lan s\mp k \ran^{-(d+1)}\Big)  \\
&\le
C_{M,n,r}^\#(L)\big(CL^{2d}\big) =  C'^\#_{M,n,r}(L).
\end{split}
\ee
\qed

\subsection{Analysis of equation \eqref{w-eq}}
\lbl{s:complete}

It this section we assume the subcritical scaling \eqref{subcr} and complete the proof of Theorem~\ref{t:quasi-exact}.
We work in the event
\be\lbl{N=}
\Om_L(M,r):=\Om_L^0(M,N,r+d+1) \qmb{with}\qu N := \Big\lceil\frac{d+
3}{\al}\Big\rceil+1,\qu r\ge 0,
\ee
where  $\Om^0_L$ is constructed in \eqref{Om-a}, \eqref{Om-cL}. 
Let us start by analysing the terms from the r.h.s. of eq. \eqref{main_eq}. 
For any $x,y,z\in\frak X_r$,
\[
\big|\cY(x,y,z)\big|_{\frak X_r}\le L^{-2d+1}\sum_{s_1,s_2,s_3,s\in\Z^d_L}\lan s\ran^r\de'^{s_1s_2}_{s_3s}|x_{s_1}|_\infty|y_{s_2}|_\infty |z_{s_3}|_\infty \,,
\]
where we recall that $|x_s|_\infty:=\sup_{\tau\in[0,T]}|x_s(\tau)|$. Then,
using that $\lan s\ran^r \le  3^r\big(\lan s_1\ran^r + \lan s_2\ran^r  + \lan s_3\ran^r\big)$ once $\de'^{s_1s_2}_{s_3s}\ne 0$, we get
\be\lbl{Y_Xr}
\big|\cY(x,y,z)\big|_{\frak X_r} \le C_rL^{d+1}|x|_{\frak X_r} |y|_{\frak X_r} |z|_{\frak X_r}\,.
\ee
In the event $\Om_L$, due to  \eqref{Om-a}, 
\be\lbl{a,A-b}
|a^{(m)}|_{\frak X_r}\le C L \qu\forall m\le M,
\qmb{so that}\qu
|A^M|_{\frak X_r}\le C_{M} L.
\ee
 Accordingly, by \eqref{Y_Xr}, 
\be\lbl{Q_in_Om} 
|\cR|_{\frak X_r} \le C_{M,r} L^{d+4} \qnd  
|\cY^{sym}(x,y, A^{M})|_{\frak X_r} \le C_{M,r} L^{d+2} |x|_{\frak X_r} |y|_{\frak X_r},
\ee 
for any $ x,y\in\frak X_r$.

Now we are aiming to invert the operator $(\Id - \eps \cL)$ from the l.h.s. of \eqref{main_eq} in the event $\Om_L(M,r)$.
\begin{lemma}\lbl{l:L_in_Om}
	In the event $\Om_L=\Om_L(M,r)$ with $M\ge 1$ 
	and $r\ge 0$ the operator 
	$(\Id - \eps \cL) :\, \frak X_{r} \mapsto \frak X_{r}$ is invertible and 
	\be\non
	\big\|(\Id - \eps\cL)^{-1}\big\|_{\frak X_r}  \le C_{M,r} L^{d+3},
	\ee
	where $\|\cdot\|_{\frak X_r}$ stands for the operator norm.
\end{lemma}
{\it Proof.} Due to representation \eqref{inv_ident}, 
to prove the existence of the inverse operator $(\Id-\eps\cL)^{-1}$ it suffices to establish the existence of the inverse
$\big(\Id - (\eps\cL)^{N}\big)^{-1}$.
Let us first show that in $\Om_L$
\be\lbl{L^n_op_bound}
\|\cL^n\|_{\frak X_r} \le C_{M,r} L^{d+3} \qquad \forall 1\le n\le N.
\ee
In the case $n=1$ this immediately follows from definition \eqref{lin_op_def} of the operator $\cL$, \eqref{Y_Xr} and \eqref{a,A-b}. Assume that $n\ge 2$.
By \eqref{L^n-deff}, for any $y\in\frak X_r$
\begin{align*}
	|\cL^n y|_{\frak X_r} &\le 
 L^{-d}\sum_{s\in\Z^d_L} \lan s\ran^r \sup_{\tau\in[0,T]} \int_0^\tau   	\sum_{\zz=\pm}\sum_{k\in\Z^d_L} |\cL_{sk}^{n,\zz}(\tau,l)||y_k(l)|\,dl 
	\\ &\le
 L^{-d}	\sum_{\zz=\pm} 	 \sum_{s,k\in\Z^d_L} \lan s\ran^r\, |y_k|_\infty\, \int_0^T \sup_{\tau\in[l,T]}  |\cL_{sk}^{n,\zz}(\tau,l)|\,dl  
	\\ &\le 
	 L^{-d+1}	\sum_{\zz=\pm} 	 \sum_{s,k\in\Z^d_L}  |y_k|_\infty\,\lan s\ran^r \lan s-\zz k\ran^{-(r+d+1)},
\end{align*}
due to \eqref{Om-cL} with $r:=r+d+1$ (see \eqref{N=}).
Note that
\[
\sum_{s\in\Z^d_L} \lan s\ran^r \lan s-\zz k\ran^{-(r+d+1)} 
	\le
	2^r\sum_{s\in\Z^d_L}\big(\lan s-\zz k \ran^r + \lan k\ran^r\big)\lan s-\zz k\ran^{-(r+d+1)} 
		\le L^d C_r\lan k\ran^r.
\]
Accordingly,
\[
	|\cL^n y|_{\frak X_r}  \le C_r L\sum_{k\in\Z^d_L}  |y_k|_\infty \lan k\ran^r = C_r L^{d+1} |y|_{\frak X_r},
\]
so \eqref{L^n_op_bound} holds.

Now, in view of \eqref{subcr}, we have
\be\lbl{L^n_norm_bound}
\|(\eps\cL)^n\|_{\frak X_r} \le C_{M,r}L^{d+3} \eps^n =  C_{M,r}L^{d+3 - \al n}\qu\qmb{for}\qu n\le N.
\ee
Since $N$ satisfies \eqref{N=}, for $L\ge L_{M,r}$ with sufficiently large $L_{M,r}$ we obtain $\|(\eps\cL)^{N}\|_{\frak X_r}\le 1/2$, so that $\|\big(\Id - (\eps\cL)^{N}\big)^{-1}\|\le 2$. Then, \eqref{inv_ident} together with \eqref{L^n_norm_bound} implies that
$
\|(\Id-\eps\cL)^{-1}\|_{\frak X_r}\le C_{M,r}  L^{d+3}.
$
\qed

\ssk

  Now  we consider a (random, defined in $\Om_L$) mapping 
  $$
F:\frak X_r\mapsto \frak X_r,\qu  F(y): =  i\eps(\Id-\eps\cL)^{-1}\Big(\cY^{sym}(y,y,A^M) + \cY(y)  + \eps^M\cR^M\Big),
  $$  
  cf. \eqref{main_eq}.
\begin{lemma}\lbl{l:contraction}
	Assume that
	\be\lbl{varrho,M}
	\varrho \ge \al^{-1}(2d+5), \qquad M\ge \al^{-1}(2d+7) + \rho.
	\ee 
	Then for any $\om\in\Om_L(M,r)$ the mapping $F=F^\om$ is a contraction of the ball $B_r(\eps^\varrho) := \{y\in\frak X_r:\, |y|_{\frak X_r} \le \eps^\varrho\}$ to itself, once $L\ge L_{M,r}$ with sufficiently large $L_{M,r}$.
\end{lemma}

{\it Proof.} 
Recall that $\eps^\rho = L^{-\al\rho}$.
Due to Lemma~\ref{l:L_in_Om} and  \eqref{Y_Xr}, \eqref{Q_in_Om}, for $\om\in\Om_L$ and $y\in B_r(\eps^\varrho)$
\be\lbl{F(y)<}
|F^\om(y)|_{\frak X_r}\le C_{M,r}L^{d+3-\al}\big(L^{ d+2-2\varrho\al} + L^{d+1-3\varrho\al} + L^{-\al M + d+4}\big).
\ee
A simple computation shows that for $\rho$ and $M$ satisfying  \eqref{varrho,M}
the r.h.s. of \eqref{F(y)<} is bounded by $C_{M,r}L^{-q}$ with $q>\varrho\al$.
Thus, $F^\om$ maps the ball $B_r(\eps^\varrho)$ to itself once $L\ge L_{M,r}$ with sufficiently large $ L_{M,r}$.

 Similarly, for any $x,y\in B_r(\eps^\varrho)$ and $\om\in\Om_L$
$$
|F^\om(x)-F^\om(y)|_{\frak X_r}\le C_{M,r} L^{d+3-\al}\big(L^{d+2-\varrho\al} + L^{d+1 -2\varrho\al}\big) |x-y|_{\frak X_r}.
$$
The r.h.s. above is bounded by  $C_{M,r}L^{2d+5 - (\rho+1)\al} |x-y|_{\frak X_r}$, 
so $F^\om$ is a contraction of the ball $B_r(\eps^\varrho)$ once $\varrho$ satisfies the first relation in \eqref{varrho,M} and $L$ is sufficiently large.
\qed

\ssk
Lemma~\ref{l:contraction} implies that for $\om\in\Om_L$ the mapping $F^\om$, restricted to the ball $B_r(\eps^\varrho)$, has a unique fixed point $w^{M,\om}$. The latter is a unique solution to eq. \eqref{main_eq}, satisfying the desired bound $|w^{M,\om}|_{\frak X_r} \le \eps^\varrho$. 


It remains to show that eq. \eqref{main_eq} has a unique solution not only in the ball $B_r(\eps^\varrho)$ but in all $\frak X_r$. Let us provide a space of sequences $v=\{v_s\in\C: \, s\in\Z^d_L\}$ with the norm
\[
|v|_{1,r} = L^{-d}\sum_{s\in\Z^d_L} \lan s\ran^r |v_s|,
\]
so that for $x\in\frak X_r$ we have $|x(\tau)|_{1,r} \le |x|_{\frak X_r}$ and $x=0$ $\Leftrightarrow$ $|x(\tau)|_{1,r}=0$ for any $0\le \tau\le T$. Literally repeating the argument used to deduce \eqref{Y_Xr}, we find
\[
|Y(x,y,z)|_{1,r} \le C_r L^{d+1}|x|_{1,r} |y|_{1,r} |z|_{1,r}.
\]
Using this estimate together with the definitions \eqref{lin_op_def} and \eqref{cYdef} of the operators $\cL$ and $\cY$, the uniqueness immediately follows by  the Gr\"onwall inequality in a standard way. Proof of Theorem~\ref{t:quasi-exact} is completed.

\subsection{Proof of Theorem~\ref{t:main}}
\lbl{s:main_t_proof}
 

\textit{(i)} We fix any  $\rho$ and $M$ satisfying assumptions of Theorem~\ref{t:quasi-exact}, and take for $\Om_L(r)$ the obtained there event $\Om_L(M,r)$.
Then the desired unique solution is given by formula \eqref{a-A-w}, in which $w_s^M$ is the unique solution from Theorem~\ref{t:quasi-exact}. 
To get the bound \eqref{b_main}, we write
\be\non
\begin{split}
\Big|\EE_{\Om_L} |a_s(\tau)|^2 - \fm(s,\tau) \Big| &\le 
\Big|\EE_{\Om_L} |a_s|^2 - \EE_{\Om_L} |A^M_s|^2  \Big|+
\Big|\EE_{\Om_L} |A^M_s|^2 - \EE |A^M_s|^2 \Big| 
\\ & + 
\Big|\EE |A^M_s|^2 - \fm(s,\tau) \Big| =:I_1+I_2+I_3.
\end{split}
\ee
Recall that $\eps=L^{-\al}$, $\al\in (0,1/2]$, so that $L\ge \eps^{-2}$. Then for $L\ge L_r$ assumptions of Theorem~\ref{t:DKMV} and Proposition~\ref{t:DKMV-extention} are fulfilled, so we have $I_3(\tau)\le C_r \lan s\ran^{-r}\eps^3$ uniformly in $\tau\ge 0$. 
Next, 
\be\non
I_2= \EE_{\Om_L^c}|A^M_s|^2 \le
\Big(\EE |A^M_s|^4\Big)^{1/2}  \Big(\PP(\Om_L^c)\Big)^{1/2}
\le C^\#(s) C_r^\#(L)\le C_r\lan s\ran^{-r}\eps^3 , 
\ee
due to bounds \eqref{bound_a} and \eqref{Om_L_prob}.
Finally, to estimate $I_1$ we note that
\[
\big||a_s|^2 - |A^M_s|^2 \big|\le
\big|w_s^{M}\big|\; \big(|w_s^M| + 2|A_s^M|\big). 
\]
In $\Om_L$ we have
\[
|w_s^M(\tau)|\le L^d|w^M|_{\frak X_r} \lan s\ran^{-r} \le  L^d\eps^\varrho \lan s \ran^{-r}\le \eps^3 \lan s \ran^{-r}
\] 
for $\tau\le T$, due to the choice of $\rho$. Since $\EE |A_s^M| \le C^\#(s)$, we obtain 
\[
I_1(\tau) \le  \eps^3 \lan s \ran^{-r} \big( \eps^3 \lan s \ran^{-r} + 2\EE |A_s^M|\big)\le C_r\eps^3 \lan s \ran^{-r},\qu \tau\le T.
\]
 Collecting the estimates above together, we get 
 \eqref{b_main}.

\textit{(ii)} We fix any $r\ge 0$ and consider the event $\Om_L(r)$ from item (1) of the theorem. 
Due to bound  \eqref{apr_bound}, the solution $a$ belongs to the space $\frak X_r$ almost surely. Then its restriction to the event $\Om_L$ a.s. coincides with the solution, constructed in item~(1), so admits approximation \eqref{b_main}. 

Next, due to the H\"older inequality, for any $p>1$ and $q=p/(p-1)$
\[
\EE |a_s(\tau)|^2 - \EE_{\Om_L} |a_s(\tau)|^2 = \EE_{\Om_L^c} |a_s(\tau)|^2 
\le 
\Big(\EE |a_s(\tau)|^{2p}\Big)^{1/p}  \Big(\PP(\Om_L^c) \Big)^{1/q}\, .
\]
Choosing $p=m/2$, in view of \eqref{apr_bound} and the estimate $\PP(\Om_L^c) \le C_r^\#(L)$, we see that this is bounded by $C^\#_r(s) L^{k/p} C_r^\#(L) \le C_r^\#(s) \eps^3$,
uniformly in $\tau\in[0,T]$.
Proof of the theorem is completed.
\qed
 
 \section{Case $d=2$}
 
\lbl{s:d=2}

The difference between cases $d=2$ and $d\ge 3$ comes from that in number-theoretic approximations borrowed from paper \cite{number_theory}.
In particular, 
for $d=2$ the r.h.s. of  \eqref{corr_nt} multiplies by $\ln L + \ln \lan(a,b)\ran$. By direct adaptation of proofs from Appendix~\ref{a:nt}, this follows from the fact that for $k=4$ the r.h.s. of \eqref{nt_smooth} multiplies by $\ln L$, see \cite[Theorem 1.4]{number_theory}.

 As it is shown in \cite{DKMV}, the "right" scaling for $d=2$ is 
\be\lbl{new_sc}
\la=\eps\nu\frac{L}{\sqrt{\ln L}}
\ee
rather than \eqref{la_sc}. 
Accordingly,  the factor $L^{-d+1}$ in eq. \eqref{a-eq}, \eqref{eff} and in definition \eqref{Ydef} of the operator $Y$   replaces by $L^{-d+1}/\sqrt{\ln L}$.

In approximation \eqref{DKMV} of Theorem~\ref{t:DKMV}, which now  holds for $L\ge\eps^{-6}$, the solution $\fm(s,\tau)$ to the WKE replaces by  $\fm(s,\tau) + f(s,\tau,L)/\ln L$, see~\cite[Remark 1.2 and Appendix D]{DKMV}. 
The correction $f(s,\tau,L)$ is uniformly bounded and has a complicated but explicit form.

Our main result,   Theorem~\ref{t:main}, stays true for $d=2$ once $\al\in (0,1/6]$ in \eqref{subcr}, \footnote{This is the assumption  $L\ge\eps^{-6}$ mentioned above which leads to the  restriction $\al\le 1/6$.} but  approximation \eqref{b_main} takes the form
\be\non
\sup\limits_{\tau\in [0,T]}	\Big|\EE_{\Om_{L}} |a_s(\tau)|^2 - \fm(s,\tau) -\frac{f(s,\tau,L)}{\ln L}\Big| \le C_{r}\lan s \ran^{-r} \eps^3, \qquad\qu \forall s\in\Z^d_L.
\ee 
The other results, formulated in Section~\ref{s:res_det}, including Theorem~\ref{t:quasi-exact}, remain unchanged.

 Because of new scaling \eqref{new_sc}, in Section~\ref{s:est_a} the factor $L^{-d+1}$ from \eqref{cum_decomp_lem} replaces by $L^{-d+1}/\sqrt{\ln L}$ and $L^{-(d-1)(p-1)}$ from \eqref{k-ind_est}~--- by $\big(L^{-(d-1)}/\sqrt{\ln L}\big)^{p-1}$. Proof of Proposition~\ref{t:sup_est_num} follows the same lines modulo a different induction hypotheses, corresponding to the new form of the bound \eqref{k-ind_est}. Note that the r.h.s. of \eqref{S_pi_bound} multiplies by $\ln L$ because of the discussed above modification in bound \eqref{corr_nt}.
 In Section~\ref{s:est_lin_op} similar changes should be made in Lemma~\ref{l:cum_decomp_L} and Proposition~\ref{t:sup_est_oper}. In particular, in the latter the factor $L^{-(d-1)(p+m-1)}$ replaces by $\big(L^{-(d-1)}/\sqrt{\ln L}\big)^{p+m-1}$. In addition, in the definition of the function $\Ga_{\cJ}$ from \eqref{GaJ}, the power $\epsilon$ should be replaced by $\epsilon/2$. This is needed to kill by the term $\ga_{\xi_1}^{-\mI_{n_1\ge 2}}$ from \eqref{gaGa} the factor $\ln\lan\xi_1\ran$, appearing in \eqref{for_ln}, again because of the modified form of \eqref{corr_nt}.
  The  r.h.s. of \eqref{bound_high_mom_num_oper} should be multiplied by $\ln\lan s\ran$,  by the same modification in \eqref{corr_nt}.
 
 Since the only difference between the cases $d=2$ and $d\ge 3$ in the bounds of Propositions~\ref{l:bound_a} and \ref{l:bound_high_mom_oper_sup} is the just mentioned factor  $\ln\lan s\ran$ in the latter and the replacement $\epsilon\rightarrow\epsilon/2$, results and proofs in Section~\ref{s:theorems} stay unchanged. One could slightly improve the employed there estimates using the factor $(\ln L)^{-1/2}$ that appears in the definition \eqref{Ydef} of the operator $Y$, but there is no reason to do that.

	\appendix
	
	\section{Wave kinetic integral}
	\lbl{a:WKI}
	
In this section define the wave kinetic integral $K(s,\tau)$ from \eqref{WKE}. Let us denote by $\Sigma_s$ the {\it resonant quadric}
	\be\non
	\Sigma_s = \{(s_1,s_2)\in\R^{2d}:\, \om^{s_1s_2}_{s_3s} = 0\},
	\ee
	where $\om^{s_1s_2}_{s_3s}$ is defined in \eqref{oms} with $s_3$ found from the equation $\de'^{s_1s_2}_{s_3s} \ne 0$ (see \eqref{def_del}), that is $s_3 := s_1+s_2-s$. Let $\mu^{\Sigma_s}$ be the measure on $\Sigma_s$ given by
	\be\non
	\mu^{\Sigma_s} (ds_1ds_2) = \big(|s_1-s|^2 + |s_2-s|^2)^{-1/2}\,ds_1ds_2\big|_{\Sigma_s},
	\ee
	where $ds_1ds_2\big|_{\Sigma_s}$ denotes the volume element on $\Sigma_s$, induced from the standard Euclidean structure on $\R^{2d}$. Then, for a function $y:\R^d\mapsto \R$ we set
	\be\lbl{i:KI}
	K(s,\tau)y
	=4C_d\int_{\Sigma_s} \mu^{\Sigma_s}(ds_1 ds_2) 
	\Bigl(\cZ^4 y_1y_2y_3 
	+
	\cZ^3y_1y_2 y_4- \cZ^2 y_1 y_3y_4
	-  \cZ^1
	y_2y_3y_4\Bigr),
	\ee
	where $y_j:=y(s_j)$ with  $s_4:=s$ and  $s_3:=s_1+s_2-s$, while $C_d$ is a number-theoretic constant satisfying $C_d\in(1, 1+2^{2-d})$.
The kernels 
	$\cZ^j=\cZ^j(\tau,\bm s)$, $\bm s = (s_1, s_2, s_3, s_4) \in   (\R^d)^4,$
are given by 
	\be\lbl{Z-edi}
\cZ^j(\tau,\bm s) := \int_0^{\tau}dl\,\e^{-\ga_{s_j}(\tau-l)} \prod_{\substack{n=1,2,3,4
		\\ n \neq j}}
\frac{\sinh(\ga_{s_n}
	l)}{\sinh(\ga_{s_n} \tau)}\,\quad \text { if } \  \tau>0,
\ee
and $\cZ^j(0,\bm s)=0$. The integral in \eqref{Z-edi} can be computed explicitly, see formula (4.15) in \cite{DKMV}. 
Let $\cC_r(\R^d)$, $r\ge 0$, denote a space of continuous functions $y:\R^d\mapsto\R$ provided with the norm
\[
|y|_{\cC_r}:=\sup\limits_{s\in\R^d} |y(s)|\lan s\ran^r.
\]
In \cite{DKMV} it is proven that for $r>d$ and $\tau\ge 0$ the operator $K(\tau)$ is a continuous mapping from $\cC_r(\R^d)$ to  $\cC_{r+1}(\R^d)$,  and for any $y\in\cC_r(\R^d)$ the curve $\tau\mapsto K(\tau)(y)$ is H\"older continuous in  $\cC_r(\R^d)$. Using this, in  \cite[Lemma 5.5]{DKMV} it is shown that  WKE \eqref{WKE} has a unique solution  $m(\tau)$ in the space of continuous $\cC_r(\R^d)$-valued functions, $\forall r>d$ once $\eps\le \eps_r$ with sufficiently small $\eps_r$.

Moreover, in \cite{DKMV} it is shown that $0\leq \cZ^j\leq 1$ and when $\tau\to\infty$ the operator $K(\tau)$ exponentially fast converges to a limiting operator $K(\infty)$,  
	given by \eqref{i:KI} with $\cZ^j$ replaced by
	$	(\ga_{s_1}+ \ga_{s_2}+\ga_{s_3}+\ga_{s_4})^{-1}$ for all $j$.
	The operator $K(\infty)$ is similar 
	to the standard four-waves kinetic operator $K_{stand}$ of WT  but still is different from the latter. Indeed, 
	$K(\infty)$ depends on the spectrum $\{\gamma_s\}$ of the damping operator $\frak A$, while $K_{stand}$ has the form \eqref{i:KI} with $\cZ^j\equiv const$ (see e.g. in \cite{Naz11}).
	Note that earlier the kinetic operator $K(\infty)$ was heuristically obtained in \cite{KM15}.

	\section{Cumulants}
	\lbl{a:cum}

	Let $v = (v_1,\dots,v_n)$ be a complex random vector such that $\EE|v_j|^k<\infty$ for every $k>0$. Consider its characteristic function 
	\[
	\psi_v(\la): = \EE e^{iv\cdot \la}, \qquad \la=(\la_1,\dots,\la_n)\in\C^n,
	\]
	where $v\cdot\la:= \sum_{j=1}^n \Re \big( v_j \bar \la_j\big)$. It is known to be a generating function for moments of the vector $v$. In particular, the moment 
$
	\cM(v):= \EE\prod_{j=1}^n v_j
$
can be computed as	
	\[
	\cM(v) = (-2i)^{n}\p_{\bar\la}\psi_v(0),
	\qmb{where}\qu \p_{\bar\la}:=\prod_{j=1}^n\p_{\bar\la_j}.
	\]
A cumulant $\ka(v)$  of the random vector $v$ is defined via the generating function $\ln\psi_v(\la)$, see \cite[Section~II.12]{Shir}. Namely,
\be\lbl{cum_gen}
\ka(v):=(-2i)^{n} \p_{\bar \la}(\ln\psi_v)(0).
\ee	
Since the moment $\cM(v)$ and cumulant $\ka(v)$  do not depend on the order of components in the vector $v$, we will view them as functions  $\cM(V)$ and $\ka(V)$ of the multi-set $V=\{v_1,\dots,v_n\}$. This well suits  for our work since we will need to apply the set operations to $V$. 

\ssk
{\it Some properties of cumulants}. We recall Notation~\ref{s:not}(2,3).

$\bullet$ The cumulant is a multi-linear function.

$\bullet$ According to definition \eqref{cum_gen},
\be\lbl{cum1,2}
\ka\big(\{v_j\}\big) = \EE v_j \qmb{and}\qu \ka\big(\{v_i,v_j\}\big) = \EE(v_iv_j) - \EE v_j \EE v_j \qquad \forall i,j.
\ee
$\bullet$ If a component $v_j$ is independent from the other components $(v_k, k\ne j)$, then $\ka(V)=0$ (e.g. this is the case when $v_j=const$). 

$\bullet$ If the vector $v$ has a multivariate normal distribution, then 
\be\lbl{cum_normal}
\ka(V) = 0 \qmb{once}\qu |V|\geq 3.
\ee

$\bullet$  Moments can be expressed via cumulants and vice verse by the Leonov-Shiryaev formula  \cite{Shir, PT}, that can be taken as an equivalent definition of cumulants: 
\be\lbl{mom-cum}
\cM(V) = \sum_{\pi\in \P(V)}\prod_{\cA\in\pi} \ka(\cA),
\qquad \ka(V) = \sum_{\pi\in \P(V)}(-1)^{|\pi|-1}(|\pi|-1)!\prod_{\cA\in\pi} \cM(\cA)\, .
\ee

$\bullet$ In our work we intensively use the following  particular case of the Malyshev formula \cite[Proposition 3.2.1]{PT}. Let $W=\{w_1,\dots,w_k\}$ be another multi-set of complex random variables satisfying $\EE|w_j|^p<\infty$ $\forall j,p$. Then 
\be\lbl{Mal}
\ka\Big(V\sqcup \big\{\prod_{i=1}^k w_i\big\}\Big) = \sum_{\pi\in \P_2(V\sqcup W)} \prod_{\cA\in\pi}\ka(\cA).
\ee

	\section{Bounds for sums over quadrics}
	\lbl{a:nt}
	
	Let us consider a non-degenerate quadratic form with integer coefficients on $\R^k$, $k\ge 5$,
	\[
	q(z)= \frac12 z\cdot Mz.
	\]
	The integer non-degenerate matrix $M$ can be chosen to be symmetric
	with even diagonal elements. 
	We consider a quadric
	\[
	\Sigma=\{z\in\R^k:\, q(z)=0\}.
	\]
	For a sufficiently fast decaying at infinity function  $f:\R^k\mapsto\R$ let us denote
	\[
	S_{L}(f)= \sum_{z\in\Z^k_L\cap \Sigma} f(z),\qquad \Z^m_L= L^{-1}\Z^m, \qu L\ge 1.
	\] 
	Our goal is to bound the sum $S_{L}(f)$.
	Assuming that the function $f$ is $C^{n_1}$-smooth, $n_1\ge 0$, we define a family of norms
	\[
	\|f\|_{n_1,n_2} = \sup_{z\in\R^k}\sup_{|\al|_1\le n_1}|\p_\al f(z)|\lan z\ran^{n_2}	,  
	\quad n_2\ge 0,
	\]
	where $\al=(\al_1,\dots, \al_k)$,  $\al_j\in\N\cup\{0\}$, is a multi-index, $|\al|_1 = \sum_{j=1}^k \al_j$ and $\p_\al = \p_1^{\al_1}\cdots \p_k^{\al_k}$. 
	In \cite{number_theory}, refining the result  by Heath-Brown \cite{HB} based on an appropriate form  of the  circle method, we found asymptotics for the sum $S_{L}$, assuming that $\|f\|_{n_1,n_2}<\infty$ for sufficiently large $n_1,n_2$. Now we establish the following corollary of this result. 
 	\begin{proposition}\lbl{l:nt}
		Assume that a function $f$ satisfies $\|f\|_{0,N(k)}<\infty$ for sufficiently large $N(k)$.
		 Then
	\be\lbl{nt_est}
	|S_{L}(f)|\le C_{M,k} L^{k-2} \|f\|_{0,N(k)}.
	\ee
	\end{proposition}
	{\it Proof}.
	If $\|f\|_{n_1,n_2}<\infty$ with sufficiently large $n_1,n_2$, depending on $k$, then 
	Theorems 1.3 and 7.3 from \cite{number_theory} immediately imply
	\footnote{In Theorem 1.3 from \cite{number_theory}  we prove that the sum $S_L(f)$ has asymptotics $L^{k-2}I(f)$ with certain $I(f)$ when $L\to\infty$ and show that $|S_L(f) - L^{k-2}I(f)|\le C_{M,k} L^{k/2+\de} \|f\|_{n_1,n_2}$, $\de>0$. In Theorem 7.3, in particular, we establish the bound $|I(f)| \le C_{M,k}\|f\|_{n_1,n_2}$. }
	 that
	\be\lbl{nt_smooth}
		|S_{L}(f)|\le C_{M,k} L^{k-2} \|f\|_{n_1,n_2}.
	\ee
	Thus, it remains to get rid of the smoothness assumption on $f$. We first
	assume that $\supp f\subset B_R:=\{|z|\le R\}$ for some $R\ge 1$.
	Let us take a $C^\infty$-smooth function $\phi:\R^k\mapsto \R_{\ge 0}$, such that $\phi|_{B_1}=1$, $\phi|_{B_2^c}=0$ and $\phi\le 1$. Then $|f(x)| \le \phi(x/R)\|f\|_\infty$ and
	\be\lbl{nt_smooth_cor}
		|S_{L}(f)|\le \|f\|_\infty S_{L}\Big(\phi\big(\frac{\cdot}{R}\big)\Big) \le C_{M,k}\|f\|_\infty L^{k-2} R^{n_2},
	\ee	 
	according to \eqref{nt_smooth}.
	
	Now we pass to a general case, when $\supp f$ is allowed to be not compact. We write  $1=\sum_{j=0}^\infty\chi_j$, where  $\chi_0=\mI_{(-\infty, 1]}$ and $\chi_j=\mI_{(2^{j-1},2^j]}$, $j\ge 1$, so that 
	\[
	f = \sum_{j=0}^\infty f_j, \qquad f_j(z) = f(z) \chi_j(|z|).
	\]
	Then for $j\ge 1$ we have $\supp f_j\subset B_{2^j}\sm B_{2^{j-1}}$, so that $\ds{\|f\|_\infty \le \frac{\|f\|_{0,N}}{2^{N(j-1)}}}$ and $\|f_0\|_\infty \le \|f\|_{0,N}$, $\forall N\ge 0$. 
	Due to \eqref{nt_smooth_cor}, this implies
	\[
	|S_{L}(f)| \le C_{M,k}L^{k-2}\Big(\|f\|_{0,N} + \sum_{j=1}^\infty \frac{\|f\|_{0,N}}{2^{(j-1)N}} 2^{jn_2}\Big) \le 
	C_{M,k,N}L^{k-2}\|f\|_{0,N},
 	\]
once $N>n_2$.
\qed	

\ssk
Let us now study the shifted sums 
\be\lbl{shift_sums}
S_{L}^{z_0}(f):= \sum_{z\in\Z^k_L\cap \Sigma} f(z-z_0), \qquad  z_0\in\R^k.
\ee
\begin{corollary}\lbl{l:nt_bound} Let the function $f$ satisfies $\|f\|_{0,N(k)}<\infty$ with $N(k)$ from Proposition~\ref{l:nt}. 
	Then
	\[
	|S_{L}^{z_0}(f)| \le C_{M,k} L^{k-2}\lan z_0\ran^{k-2}\|f\|_{0,N(k)}.
	\]
	\end{corollary}
	{\it Proof.} Denoting 
	$u:=z/\lan z_0\ran$, we write
	\[
	S_{L}^{z_0}(f) = \sum_{u\in\Z^k_{L\lan z_0\ran}\cap \Sigma} f\big(\lan z_0\ran u - z_0\big).
	\]
	Due to  Proposition~\ref{l:nt}, 
	\[
	|S_L^{z_0}(f)| \le C_{M,k}(L\lan z_0\ran)^{k-2} \|g^{z_0}\|_{0,N(k)},
	\]
	where $g^{z_0}(u) = f\big(\lan z_0\ran u - z_0\big)$.
Using the triangle inequality, we find
\begin{align*}
\|g^{z_0}\|_{0,N(k)} &= \sup_{u\in\R^k}\lan u\ran^{N(k)}|g^{z_0}(u)| = \sup_{z\in\R^k}\Big\lan \frac{z}{\lan z_0\ran} \Big\ran^{N(k)}|f(z-z_0)|\\
& \le \sup_{z\in\R^k} \Big(\Big\lan \frac{z-z_0}{\lan z_0\ran} \Big\ran^{N(k)}+ 1  \Big)|f(z-z_0)|
\le 2\|f\|_{0, N(k)}.
\end{align*}
\qed

We will apply this corollary in the case when $z=(u,v)\in\R^{d}\times\R^d$, $d\ge 3$, $q(z) = u\cdot v$ and  $f(z) = \lan z\ran^{-\mu}$. 
\begin{corollary}\lbl{l:corr_nt} If  $\mu\ge \mu_d$ with sufficiently large $\mu_d$ then for any $a,b\in\R^d$
\be\lbl{corr_nt}
\sum_{u,v\in\Z^d_L:\, u\cdot v=0} \frac{1}{\big\lan (u+a,v+b)\big\ran^\mu} \le C_d L^{2(d-1)} \, \big\lan(a,b)\big\ran^{2(d-1)} \, .
\ee
\end{corollary}

\begin{conjecture}\lbl{conj:nt}
	The shifted sum \eqref{shift_sums} satisfies estimate \eqref{nt_est} uniformly in $z_0$. Accordingly, the sum from \eqref{corr_nt} is bounded by $C_d L^{2(d-1)}$ uniformly in $a,b$.
\end{conjecture}

	\section{Several proofs}
	\lbl{a:sev_proofs}

	\subsection{Proof of Lemma~\ref{l:cum_basics_L}}
	\lbl{a:cum_prop_op}
	
To prove items (1) and (2) we argue by induction in $\cN:=\sum_{j=1}^m n_j$. Assume that $\cN=0$. If additionally $m=0$, the assertion follows from Lemma~\ref{l:cum_basics}. 
	If $m\ge 1$ then, due to \eqref{LW^0}, $X^j$ are either constants $\de_{\xi_j\s_j}$ or zeros, so $\ka(\cJ)$ may not vanish only if $m=1$, $k = 0$ and $X^1_{\xi_1\s_1} =  \de_{\xi_j\s_j}$ (see Appendix~\ref{a:cum}). In this case $I_+(\cJ) = \{\xi_1\}$ and $I_-(\cJ) = \{\s_1\}$, so the assertions hold.
	
	Let items (1) and (2) be proved for  some $\cN\ge 0$. To establish them for $\cN+1$, we assume for definiteness \eqref{for_definit} and use \eqref{ka_J-est_L} (or equivalently \eqref{cum_decomp_lem_L}).  
	By the induction hypotheses, each term $\ka(\cA)$ in \eqref{ka_J-est_L}  vanishes unless 
	$|I_+(\cA)| = |I_-(\cA)|$ and $\sum I_+(\cA) = \sum I_-(\cA)$.
	At the same time, as in the proof of Lemma~\ref{l:cum_basics}, it is straightforward to check that  for any partition $\pi$ in \eqref{ka_J-est_L}
	\[
	\bigsqcup_{\cA\in\pi} I_+(\cA) = \{s_1,s_2\}\sqcup I_+(\cJ)\sm\xi_1
	\qnd 
	\bigsqcup_{\cA\in\pi} I_-(\cA) = \{s_3\}\sqcup I_-(\cJ).
	\]
	Then, taking into account the identity $s_1+s_2-\xi_1=s_3$ (provided by the factor $\Del^{s_1s_2}_{s_3\xi_1}$ in \eqref{ka_J-est_L}), we get the equality from items (1,2) of the lemma.
	
	Item (3) directly follows from item (1). Indeed, 
	\[
	I_\pm(\cJ) = \bigsqcup_{j=1}^m I_\pm\big(\{X^j_{\xi_j\s_j}\}\big) \sqcup I_\pm\big(\{\boldsymbol{x}_{\bbeta} \}\big),
	\]
	where $I_\pm\big(\{\boldsymbol{x}_{\bbeta}\}\big):= I_\pm\big(\{x_{\eta_1}^1,\dots,x_{\eta_k}^k\}\big)$
	Since $|I_+\big(\{X^j_{\xi_j\s_j}\}\big)| - |I_-\big(\{X^j_{\xi_j\s_j}\}\big)| $ is even for every $j$, item (1) implies that
	$|I_+\{\boldsymbol{x}_{\bbeta}\}| - |I_-\{\boldsymbol{x}_{\bbeta}\}|$ also is. Accordingly, $k = |I_+\{\boldsymbol{x}_{\bbeta}\}| + |I_-\{\boldsymbol{x}_{\bbeta}\}|$ is even as well.
	\qed

	\subsection{Proof of bound \eqref{des_est_lin}: other cases}
	\lbl{a:S_pi}
	
In this section we conclude the proof of estimate \eqref{des_est_lin} in the case $|\pi|=2$, so that $\pi=\{\cA_1,\cA_2\}\in\cP_2(\check \cJ\sqcup \cJ_{new}^\zz)$. 

$\bullet$ 
Assume that elements of the set $\cJ_{new}^\zz$ with the indices $s_1, s_2$ belong to $\cA_1$ while that with the index $s_3$ lies in $\cA_2$ (e.g. for $\zz=+$ this means that $A_{s_1}, \cL^{n_1-1,\zz_1}_{s_2\s_1}\in\cA_1$ and $\bar A_{s_3}\in\cA_2$).
Then, by \eqref{frak_S_L}, $s_3 = s_3(\bxi,\bsi,\bbeta;\pi)$ is defined uniquely once $\bm s\in\frak S(\bxi,\bsi,\bbeta;\pi)$.  Accordingly,  in \eqref{S_pi_bound_L} vector $u=s_3-\xi_1 - v$ is fixed, so
the summation is performed only over  $v$ satisfying
\[
u\cdot v=(s_3 - \xi_1 - v)\cdot v = 0.
\]
Then $w:=v-(s_3-\xi_1)/2 \in \Z^d_{2L}$ satisfies
\be\lbl{w_sp}
|w|^2 - \Big|\frac{s_3-\xi_1}{2}\Big|^2=0.
\ee
Note that
\begin{align}\non
	C^\#(u+\xi_1,v + \xi_1\de_{\zz\, -}) &= C^\#(s_3 -v, v +  \xi_1\de_{\zz\, -}) \\\non
	&\le  C_1^\#(s_3 + \xi_1\de_{\zz\, -}) C_1^\#(2v + \xi_1\de_{\zz\,-}-s_3)
	\\\lbl{CcC=}
	 &=
	 C_1^\#(s_3 + \xi_1\de_{\zz\, -}) C_1^\#(2w - \xi_1\de_{\zz\,+})\,.
\end{align}	
 In the case $\zz=-$ the latter product is bounded by $C^\#(2w) \le C_1^\#(s_3-\xi_1)$, by \eqref{w_sp}.
 Since the number of points $w\in \Z^d_{2L}$ on the sphere \eqref{w_sp} is bounded by $C_d L^{d-1} |s_3-\xi_1|^{d-1}$, we get
\be\non
S_\pi^- \le C^\#(\xi_1-\zz_1\s_1)C^\#(s_3-\xi_1)|s_3-\xi_1|^{d-1}L^{d-1} \le C_1^\#(\xi_1-\zz_1\s_1 ) L^{d-1}\, ,
\ee
which is in accordance with \eqref{des_est_lin}.
In the case $\zz=+$ the r.h.s. of \eqref{CcC=} is bounded by $C^\#(s_3)$, so
\be\non
S_\pi^+ \le C^\#(\xi_1-\zz_1\s_1)C^\#(s_3)|s_3-\xi_1|^{d-1}L^{d-1} \le C_1^\#(\xi_1-\zz_1\s_1 )\lan \xi_1\ran^{d-1} L^{d-1}\, ,
\ee
which is in accordance with \eqref{des_est_lin} if $n_1\ge 2$. 
If $n_1=1$ then $\cL^{n_1-1,\zz_1}_{s_2\s_1}\in\cA_1$ is not random but $|\cA_1|\ge 2$, so  $\ka(\cA_1)=0$. Accordingly, $\frak S(\bxi,\bsi,\bbeta;\pi)=\emptyset$, so that $S_\pi^+=0$.

$\bullet$ It remains to consider the case when elements of the set $\cJ_{new}^\zz$ with the indices $s_1, s_3$ are in $\cA_1$ while that with the index $s_2$ belongs to $\cA_2$.
Then  $s_2$ is defined uniquely for $\bm s\in\frak S(\bxi,\bsi,\bbeta;\pi)$.  So, $v=s_2-\xi_1$ is fixed, and
the summation in \eqref{S_pi_bound_L} is performed only over vectors $u$ from the hyperplane 
\[
u\cdot v=u\cdot (s_2 - \xi_1)= 0.
\]
Then, $S^\zz_\pi   
\le  C^{\#}(\xi_1-\zz_1 \s_1)L^{d-1}$ if $s_2\ne\xi_1$. Otherwise, due to the identity $\de'^{s_1s_2}_{s_3\xi_1}=1$, we have $s_1=s_2=s_3=\xi_1$, so the summation in \eqref{lin_S_pi}, \eqref{lin_S_pi'} is absent and we again get the desired bound.
\qed

\subsection{Proof of estimate \eqref{W_est}}
\lbl{a:W_est}
Let $\cJ$ be as in \eqref{JJ} except that $X^1_{\xi_1\s_1}=\cW^{n_1,\zz_1}_{\xi_1\s_1}(\tau_1,t_1)$, $n_1\ge 2.$
Writing $\cW^{n_1,\zz_1}_{\xi_1\s_1}$ via its definition \eqref{W_def}, we see that $\ka(\cJ) = i\cK(\tau_1)$, where $\cK(l)$ is given by \eqref{KKK}. Then, by \eqref{lin_op_K}, we get \eqref{ka_J-est_L} in which the factor $\ga_{\xi_1}^{-\mI_{n_1\ge 2}}$ is removed. Then, 
assertions of Lemmas~\ref{l:cum_basics_L} and \ref{l:cum_decomp_L} are satisfied, but without this factor in decomposition  \eqref{cum_decomp_lem_L}. Since the sets $\cA$ from the r.h.s. of \eqref{cum_decomp_lem_L} are of the form \eqref{JJ}, the corresponding cumulants $\ka(\cA)$  satisfy estimate \eqref{sup_est_num_L}. Then, \eqref{cum_decomp_lem_L} implies  \eqref{k(J)<1}, again without the factor  $\ga_{\xi_1}^{-\mI_{n_1\ge 2}}$, where $S_\pi^\zz$ satisfies \eqref{des_est_lin}.
Accordingly, we obtain
\be\lbl{kakaka}
	|\ka(\cJ)| \le  \lan\xi_1\ran^{2(d-1)}\, C^\#(\bxi - \zz\odot\bsi) C^\#(\bbeta)  L^{-(d-1)(p + m -1)},
\ee 
where we used the bound $\Ga_{\check\cJ\sqcup\cJ^{\zz}_{new}}^{-1} \le 1.$

Let now $\cJ$ be as in \eqref{JJ} except that $X^1_{\xi_1\s_1}=\cW^{n_1,\zz_1}_{\xi_1\s_1}(\tau_1,t_1)$ and $X^2_{\xi_2\s_2}=\ov{\cW^{n_2,\zz_2}_{\xi_2\s_2}}(\tau_2,t_2)$, where $n_1, n_2\ge 2$.
We again write $\cW^{n_1,\zz_1}_{\xi_1\s_1}$ via its definition \eqref{W_def} and get \eqref{cum_decomp_lem_L} without the factor $\ga_{\xi_1}^{-\mI_{n_1\ge 2}}$, where for every partition $\pi$ one of the sets $\cA\in\pi$ contains the term $\ov{\cW^{n_2,\zz_2}_{\xi_2\s_2}}$, while the other are of the form \eqref{JJ}. So, one of the cumulants $\ka(\cA)$ from \eqref{cum_decomp_lem_L} satisfies estimate \eqref{kakaka} with $\lan\xi_1\ran$ replaced by $\lan\xi_2\ran$, while the other satisfy \eqref{sup_est_num_L}. Accordingly, as in \eqref{k(J)<1},  \eqref{des_est_lin} we get 
\be\lbl{kaWW}
|\ka(\cJ)| \le  \Big(\lan\xi_1\ran\lan\xi_2\ran\Big)^{2(d-1)}\, C^\#(\bxi - \zz\odot\bsi) C^\#(\bbeta)  L^{-(d-1)(p + m -1)}.
\ee 
Now to obtain \eqref{W_est} it remains to use in \eqref{ka-2mom} bounds \eqref{kakaka} with $\cJ=\big\{\cW^{n,\pm}_{sk}\big\}$ and \eqref{kaWW} with $\cJ=\big\{\cW^{n,\pm}_{sk},\ov{\cW^{n,\pm}_{sk}}\big\}$.
\qed

\section{
	Control of constants in Proposition~\ref{l:bound_a}}
\lbl{a:const_growth}

In this section we prove
\begin{proposition} \lbl{l:bound_a_a} 
	For any $m\ge 0$ and $q\ge 1$  
	\begin{align}\non
		\EE\, \big|a_s^{(m)}(\tau)\big|^{2q} \leq 	C_r\lan s\ran^{-2qr}\big(q(2m+1)\big)! \qquad \forall r\ge 0\, , 
	\end{align}
	uniformly in $\tau\ge 0$ and $L\ge 1$.
\end{proposition}
Bound \eqref{contr_const_intro} can be deduced from Proposition~\ref{l:bound_a_a} by a standard argument similar to that used 
to get Corollary~\ref{l:bound_high_mom_num}.
\smallskip

For a multi-vector $\bm\eta=(\eta_1,\dots,\eta_k)\in (\Z_L^d)^k$, $\eta_j\in\Z_L^d$, let us denote 
$$
[\bm\eta] = [\eta_1,\dots,\eta_k]:=\prod_{j=1}^k \lan \eta_j \ran, \qquad k\ge 1.
$$
The function $ \bm\eta \mapsto [\bm\eta]$ is multiplicative: for multi-vectors $\bm\eta\in(\Z^d_L)^k$ and $\bm\zeta\in(\Z^d_L)^m$,
\be\lbl{weight_mult}
[(\bm\eta,\bm\zeta)] = [\eta_1,\dots,\eta_k, \zeta_1,\dots,\zeta_m] = [\bm\eta][\bm\zeta].
\ee  
Let us consider the multi-set $J$ from \eqref{J_set}. We will view the cumulant $\ka(J)$ as a function $f = f(\bm\eta, \bm\tau)$ of the multi-vector of indices $\bm\eta$ and vector of times $\bm\tau$. We will work with the family of weighted 
$L_\infty$-norms
\be\lbl{mu-norm}
|f|_\mu:= \sup_{\bm\eta,\bm\tau}\, [\bm\eta]^{\mu}\, |f(\bm\eta, \bm\tau)|, \qquad \mu\ge 0.
\ee
If there are two functions $f$ and $g$ such that $f(\bm\eta,\bm\tau) = f\big((\eta_j)_{j\in I},\bm\tau\big)$ 
and $g(\bm\eta,\bm\tau) = g\big((\eta_j)_{j\notin I},\bm\tau\big)$ for some $I\subset\{1,\dots,k\}$,
then, in view of \eqref{weight_mult},
\be\lbl{mu-norm_mult}
|fg|_\mu \le |f|_{\mu}|g|_{\mu}.
\ee
\begin{lemma}[Induction step]\lbl{l:cum_exp_sup_num_a}
	There are constants $C_d, \,\mu_d>0$ such that for any multi-set $J$ as in \eqref{J_set}, in which $x^1$ satisfies \eqref{x^1=} with $\Theta = \cY$, 
	\be\lbl{k-ind_est_a}
	|\ka(J)|_\mu \leq C_d 3^\mu \sum_{\pi\in\P_2(\check J\sqcup J_{new})} L^{(d-1)(2-|\pi|)} \prod_{\cA\in\pi}|\ka(\cA)|_{\mu}, \qquad \forall\mu\ge \mu_d
	\ee
	and $L\ge 1$, where the multi-sets $\check J$ and $J_{new}$  are defined in \eqref{JJ_new}.
\end{lemma} 
{\it Proof.} Let us multiply \eqref{cum_decomp_lem} by $[\bbeta]^\mu$. Then, multiplying and dividing the r.h.s. of the obtained relation by $[\bm s]^\mu=[s_1,s_2,s_3]^\mu$ and using \eqref{weight_mult}, we find
\be\lbl{K_m-ind_st_a}
\begin{split}
				[\bm\eta]^\mu	\, |\ka(J)| & \le \ga_{\eta_1}^{-1} \sum_{\pi\in\P_2(\check J\sqcup J_{new})}
L^{-d+1}	\sum_{\bm s\in \frak S(\bbeta;\pi)} 
	\frac{\lan\eta_1\ran^\mu}{\big[s_1,s_2,s_3\big]^\mu} \\
	&\sup_{0\le l\le \tau_1}\prod_{\cA\in\pi}	\Big( |\ka(\cA)|\,[\bm\eta_\cA ]^\mu\, \big[\boldsymbol{s}_\cA\big]^\mu\Big)\, ,
\end{split}
\ee
where  $\bm\eta_{\cA}$ and $\bm{s}_\cA$ denote vectors of indices $\eta_i$ and $s_j$ of elements from the set $\cA$.
Accordingly,
\be\lbl{K_2*_a}
|\ka(J)|_\mu \le \ga_{\eta_1}^{-1} \sum_{\pi\in\P_2(\check J\sqcup J_{new})} 	 L^{-d+1} \,S_\pi(\bbeta) \,\prod_{\cA\in\pi} |\ka(\cA)|_\mu\,, 
\ee
where 
\be\lbl{S_pi_a}
S_\pi(\bbeta) =\sum_{\bm s\in \frak S(\bbeta;\pi)} \frac{\lan\eta_1\ran^\mu}{[s_1,s_2,s_3]^\mu}\,.
\ee
Then, it remains to show that for any $\pi\in\P_2(\check J\sqcup J_{new})$, 
\be\lbl{S_pi-est_a}
S_\pi(\bbeta) \le
C_d 3^\mu   L^{(d-1)(3-|\pi|)}\, \lan \eta_1 \ran^{2(d-1)}\, ,
\ee
since $\ga_{\eta_1}^{-1}\lan \eta_1 \ran^{2(d-1)} \le C$.
Since $\de'^{s_1s_2}_{s_3\eta_1}\ne 0$ for $\bm s\in\frak S(\bbeta;\pi)$,  
\be\lbl{sum_s-eta}
\lan\eta_1\ran^\mu\le 3^{\mu-1}\big(\lan s_1\ran^\mu + \lan s_2\ran^\mu + \lan s_3\ran^\mu\big). 
\ee
Then, making substitution \eqref{subst} and recalling \eqref{om_subst}, we get
\be\lbl{S-uv}
S_\pi(\bm\eta)\le 
3^{\mu-1}\sum_{u,v\in\Z^{d}_L:\, u\cdot v=0}  \Big( 
\frac{1}{[s_2,s_3]^\mu}
+\frac{1}{[s_1,s_3]^\mu}
+\frac{1}{[s_1,s_2]^\mu}\Big)\, ,
\ee
where $s_1=u+\eta_1$, $s_2= v+\eta_1$ and $s_3 = u+v+\eta_1$.
Since 
$[s_i,s_j]\ge \lan(s_i,s_j)\ran/2$ and $\lan a+b,b\ran \ge C^{-1}\lan a,b\ran$ for any $a,b$, 
Corollary~\ref{l:corr_nt} implies
\be\lbl{K_bar-m^1_num}
S_\pi (\bm\eta)\le  C_d3^\mu L^{2(d-1)} \lan \eta_1\ran^{2(d-1)},
\ee
once $\mu\ge\mu_d$ is sufficiently large. This is in accordance with \eqref{S_pi-est_a} if $|\pi| = 1$. 

$\bullet$ Since $\pi\in\cP_2(\check J \sqcup J_{new})$, we have either $|\pi|=1$ or $|\pi|=2,3$.
Let $|\pi|=2$, so $\pi = \{\cA_1, \cA_2\}$.  
Since $\cA_j \cap J_{new} \ne\emptyset$, $j=1,2$, for definiteness we
assume that
$\cA_1\ni y^1_{s_1}$ and $\cA_2 \ni y^2_{s_2},  \bar y^3_{s_3}$; the other cases are similar.
Then, by the second relation in \eqref{frak_S}, $s_1 = s_1(\bbeta;\pi)$ is uniquely defined once $\bm s\in\frak S(\bbeta;\pi)$, so the summation in \eqref{S-uv} is performed only over $v$ from the hyperplane $(s_1(\bbeta;\pi)- \eta_1)\cdot v = 0$. If $s_1\ne \eta_1$, we get
\be\lbl{K_bar-m^2_num_a}
S_\pi   
\le C_d 3^\mu L^{d-1} \, ,
\ee
in accordance with \eqref{S_pi-est_a} for $|\pi|=2$.
If $s_1=\eta_1$, due to the relation $\de'^{s_1s_2}_{s_3\eta_1}= 1$, we get $s_1=s_2=s_3=\eta_1$, so there is no summation in \eqref{S_pi_a} and $S_\pi\le 1$.

$\bullet$ If $|\pi| = 3$, we have
$\pi=\{\cA_1,\cA_2,\cA_3\}$ with $|\cA_k\cap J_{new}|=1$, so the set $\frak S(\bbeta;\pi)$ consists of at most one point $\bm s(\bbeta;\pi)$. Then, due to \eqref{S_pi_a}   and \eqref{sum_s-eta}, we get  $S_\pi\le 3^{\mu}$, which is again in agreements with \eqref{S_pi-est_a}.
\qed

\smallskip

Let $\cF^{(n)}_\cY \subset \cF^{(n)}$ be a set of random fields, constructed as the set $\cF^{(n)}$ in \eqref{cFn} but with $\Theta=\cY$. That is, $\cF^{(0)}_\cY = \cF^{(0)}$ and for $n\ge 1$
$$
\cF_\cY^{(n)} := \Big\{\cY(x^1,x^2,x^3):\; x^j\in\cF_\cY^{(n_j)}, \; n_1+n_2+n_3 = n-1\Big\}.
$$
We set $\cF_\cY= \bigcup_{n=0}^\infty \cF^{(n)}_\cY$.
\begin{proposition}\lbl{t:sup_est_num_a}
	For any set $J$ as in \eqref{J_set}, \eqref{ord_even} with $x^j\in\cF_\cY\cup\bar\cF_\cY$ $\forall j$,
	\be\lbl{sup_est_num_a}
	\sum\limits_{\pi\in\P(J)} L^{(d-1)(p-|\pi|)}\prod_{\cA\in\pi}|\ka(\cA)|_\mu 
	\leq \cC_{\deg J}(\mu, p),  \qquad \forall\mu\ge\mu_d,
	\ee
	where $L\ge 1$,
	\be\lbl{K_muu_a}
	\cC_N(\mu,p) = (C_d3^\mu)^{N}  (\cB_\mu)^{p+N}  (p+N)! , \quad \cB_\mu: = \sup_{s\in\R^d}\frac{\lan s\ran^{2\mu}b(s)^2}{\ga_{s}},
	\ee
	and the constants $C_d,\mu_d>0$ are from Lemma~\ref{l:cum_exp_sup_num_a}.
	Moreover,  the summands in \eqref{sup_est_num_a} with $p-|\pi|<0$ vanish.
\end{proposition}
Note that 
\be\lbl{CCC_N}
\cC_N(\mu,p)  \le C_{d,\mu}\, (p+N)! \, .
\ee
{\it Proof of Proposition~\ref{t:sup_est_num_a}}.
If $|\pi|>p=|J|/2$ then there is a multi-set $\cA\in\pi$ with $|\cA|=1$. 
So, $\ka(\cA)= 0$ and the corresponding to $\pi$ summand in \eqref{sup_est_num_a} vanishes.

To prove the estimate \eqref{K_muu_a} we argue by induction in $\deg J$. We first check the assertion  when $\deg J=0$, so that $x^j \in \{a^{(0)}, \bar a^{(0)}\}$ $\forall j$. 
In this case, due to \eqref{cum_normal} and \eqref{a^0-corr}, it suffices to take the summation in \eqref{sup_est_num_a} over the partitions $\pi$, pairing in $J$ the non-conjugated variables $a^{(0)}$ with the conjugated variables $\bar a^{(0)}$. The size $|\pi|$ of such partitions $\pi$ equals $p$ and their number is $p!$. Then, denoting the l.h.s. from \eqref{sup_est_num_a} by $\cK(J)$, due to \eqref{a^0-corr} we get
\[
\cK(J) = p! \Big|\ka\big(\{a^{(0)}, \bar a^{(0)}\}\big)\Big|_\mu^p = 
p!\Big(\sup_{s\in\Z^d_L,\; \tau\ge 0} \lan s\ran^{2\mu}\EE |a_s^{(0)}(\tau)|^2\Big)^p \le
p! (\cB_\mu)^p.
\]
Now we assume that the statement is proven for any $J$ with $\deg J = N$ and establish it for $J$ with  $\deg J = N+1$. Assume for definiteness that $x^1\in\cF_\cY$ and $\deg x^1>0$, so that \eqref{x^1=} holds with $\Theta = \cY$.  Denote by 
$\cA_1$ the element of the partition $\pi$ containing $x^1_{\eta_1}$. Then, applying Lemma~\ref{l:cum_exp_sup_num_a} to the cumulant $\ka(\cA_1)$, we get
\begin{align}\non
	\cK(J) \le   &\sum_{\pi\in\P(J)} L^{(d-1)(p-|\pi|)} \prod_{\cA\in\pi\sm \cA_1} |\ka(\cA)|_\mu 
	\\ \lbl{I^N+1=_a}
	& C_d3^\mu \sum_{\pi'\in\P_2(\check \cA_1\sqcup J_{new})} L^{(d-1)(2-|\pi'|)} \prod_{\cA'\in\pi'}|\ka(\cA')|_{\mu},
\end{align}
where  $\check \cA_1 = \cA_1\sm\{x^1_{\eta_1}\}\ne\emptyset$
and, as usual, $J_{new} = \{y^1_{s_1}, y^2_{s_2}, \bar y^3_{s_3}  \}$.
Denote also $\check J:=J\sm \{x_{\eta_1}^1\}$.
Note that 
\be\lbl{pi''pi'}
\pi'':= (\pi\sm \{\cA_1\})\sqcup \pi'
\ee
 is a partition of the set $\check J\sqcup J_{new}$.
The other way round, any partition $\pi''\in \P(\check J\sqcup J_{new})$
can be uniquely represented in the form \eqref{pi''pi'} 
with $\pi\in\cP(J)$ and $\pi'\in  \P_2(\check \cA_1\sqcup J_{new})$
where $\cA_1$, as before, is the elements of $\pi$ that contains $x^1_{\eta_1}$. Indeed, the partition $\pi'$ must consist of all  $\cA\in\pi''$ satisfying $\cA\cap J_{new}\ne \emptyset$. Then $\cA_1 = \Big(\bigsqcup_{\cA'\in\pi'}\cA'\sm J_{new}\Big) \sqcup\{x^1_{\eta_1}\}$ and $\pi = (\pi''\sm\pi')\sqcup\{\cA_1\}$.

Accordingly, since
$
|\pi''|= ( |\pi|-1) + |\pi'|,  
$
\eqref{I^N+1=_a} takes the form
\be\non
\cK(J) \le  
C_d3^\mu
\sum_{\pi''\in\P(\check J \sqcup J_{new})} L^{(d-1)(p+ 1 - |\pi''|)} \prod_{\cA\in\pi''} |\ka(\cA)|_\mu. 
\ee
Since $\deg(\check J\sqcup J_{new}) = \deg J - 1 =  N$ and  $|\check J\sqcup J_{new}| = 2p-1+3 = 2(p+1)$, by the induction hypotheses we get 
\[
\cK(J)\le C_d3^\mu \cC_{N}(\mu,p+1) = \cC_{N+1}(\mu,p).
\]
\qed

\begin{corollary} \lbl{l:bound_high_mom_num_a} 
	For any $x\in\cF_\cY\cup\bar\cF_\cY$, $p\ge 1$ and $\mu\ge \mu_d$, 
	\begin{align}\non
		\EE  |x_s(\tau)|^{2p} \leq \cC_{N}(\mu, p)\lan s \ran^{-2p\mu} \qmb{uniformly in}\qu\tau\ge 0
	\end{align}
	and  $L\ge 1$, where  $N:= 2p\deg x$ and $\cC_{N}(\mu,p)$ is the constant from  Proposition~\ref{t:sup_est_num_a}.
\end{corollary}
{\it Proof.}
We start as when proving Corollary~\ref{l:bound_high_mom_num}  and get \eqref{Ex-cumJ}.
Then, by
\eqref{mu-norm_mult},  
$$
\EE |x_s(\tau)|^{2p} \lan s \ran^{2p\mu} 
\le  \sum_{\pi\in\P(J)} \prod_{\cA\in\pi} |\ka(\cA)|_\mu \le \cC_{\deg J}(\mu, p),
$$
due to Proposition~\ref{t:sup_est_num_a}. Since $\deg J = 2 p  \deg x$, we get the assertion. 
\qed

{\bf Proof of Proposition \ref{l:bound_a_a}.}
Note that $a^{(m)}$ can be written in the form \eqref{a---x} with $x(\cT)\in\cF_\cY^{(m)}$ and that the number of ternary trees $|\cT_m|$ with $m$ internal nodes satisfies $|\cT_m|\le C^m$. Then, by Corollary~\ref{l:bound_high_mom_num_a},
\[	
	\EE\, \big|a_s^{(m)}(\tau)\big|^{2p} \lan s\ran^{2p\mu}\leq C^{2pm}\cC_{2pm}(\mu,p) \le	C_{d,\mu} \big(p(2m+1)\big)!\, , 
\]
due to \eqref{CCC_N}.
\qed

	\bsk
	
{\bf Funding.} The work on Sections 1-3 was supported by the
Ministry of Science and Higher Education of the Russian Federation (megagrant No. 075-15-2022-1115). The work   on Sections 4-E was supported by the
Russian Science Foundation under Grant no.23-11-00150.

\end{document}